\documentclass[%
  reprint, onecolumn,
superscriptaddress,
showkeys,
preprintnumbers,
 amsmath,amssymb,
 aps,
prb,
floatfix
]{revtex4-2}

\usepackage{graphicx}
\usepackage{dcolumn}
\usepackage{bm}
\usepackage{placeins}
\usepackage[center]{caption}
\usepackage{color}
\usepackage{xcolor}
\usepackage{soul}

\usepackage[pdftex,breaklinks,colorlinks,
citecolor=blue,
urlcolor=blue,
pdfsubject={LaTeX}]{hyperref}

\makeatletter
\def\maketitle{
\@author@finish
\title@column\titleblock@produce
\suppressfloats[t]}
\makeatother
\usepackage{comment}

\begin{document}

\preprint{Accepted copy of paper published in Physical Review Materials. \url{https://doi.org/10.1103/PhysRevMaterials.8.083605}}

\title{Vacancy-mediated transport and segregation tendencies of solutes in FCC nickel under diffusional creep: A density functional theory study}

\author{Shehab Shousha}

\affiliation{Idaho National Laboratory, Idaho Falls, 83415, ID, United States}
\affiliation{North Carolina State University, Raleigh, 27695, NC, United States}
\author{Sourabh Bhagwan Kadambi}%
\email[]{SourabhBhagwan.Kadambi@inl.gov}
\affiliation{Idaho National Laboratory, Idaho Falls, 83415, ID, United States}
\author{Benjamin Beeler}
\affiliation{Idaho National Laboratory, Idaho Falls, 83415, ID, United States}
\affiliation{North Carolina State University, Raleigh, 27695, NC, United States}
\author{Boopathy Kombaiah}
\affiliation{Idaho National Laboratory, Idaho Falls, 83415, ID, United States}

\date{\today}

\begin{abstract}
The Nabarro--Herring (N-H) diffusional creep theory postulates the vacancy-mediated transport of atoms under a stress gradient as the creep mechanism under low-stress and high-temperature conditions. In multicomponent alloys, we premise that this stress-assisted flow of vacancies to and from grain boundaries will produce elemental segregation. An observation of such segregation, validated with theoretical predictions, can provide the necessary experimental evidence for the occurrence of N-H creep. Theoretical calculations of the segregation tendencies via analyzing the dominant solute diffusion mechanisms and the difference in diffusivities of the elements are therefore essential. To this end, this study applies density functional theory calculations of migration barriers and solute-vacancy binding energies as input to the self-consistent mean field theory to assess the vacancy-mediated diffusion mechanisms, transport coefficients, and segregation tendencies of Co, Cr, Mo, Re, Ta, and W solutes in face-centered cubic Ni. We find Co, Re, and W to be slow diffusers at high temperatures and Cr, Mo, and Ta to be fast diffusers. Further analysis shows that the slow diffusers tend to always enrich at vacancy sinks over a wide range of temperatures. In contrast, the fast diffusers show a transition from depletion to enrichment as the temperature lowers. Furthermore, our analysis of the segregation tendencies under tensile hydrostatic strains shows that slow diffusers are largely unaffected by the strain and favor enrichment. On the other hand, the fast diffusers exhibit high sensitivity to strain and their segregation tendency can transition from depletion to enrichment at a given temperature. 
The transport coefficients calculated in this work are expected to serve as input to mesoscale microstructure models to provide a more rigorous assessment of solute segregation under N-H creep conditions.
\end{abstract}


\maketitle

\section{Introduction}

The diffusion of atoms and point defects fundamentally enables various material processes, such as creep deformation \cite{RN370}, radiation-induced microstructure evolution \cite{was_fundamentals_2016}, phase transformation \cite{easterling_phase_2009}, self-assembly \cite{RN470, RN469}, and electromigration \cite{RN470}. Therefore, an understanding of the diffusion mechanisms is essential to control and predict the processes relevant to various engineering applications. For instance, coupled diffusion kinetics of atoms and vacancies is presumed to control the rate of creep deformation in materials under low stresses at high temperatures. The governing mechanisms are referred to as diffusional creep and derive significant consideration in designing creep-resistant materials and estimating the service life of engineering components in energy applications \cite{murty_creep_1999, kassner_fundamentals_2015}. 

Seventy-five years ago, Nabarro and Herring (N-H) first theorized diffusional creep mechanisms based on stress-induced vacancy flow in materials \cite{RN432, RN433,RN475}. The theory premised that a vacancy concentration gradient established between grain boundaries (GBs) subjected to different stresses results in a vacancy flux, giving rise to plastic strain in the grain. The vacancy flux can dominate via lattice (N-H creep) or GBs (Coble creep) depending on the applied stress, temperatures, and material microstructure. The diffusional creep theories (N-H and Coble) provide a compelling picture for materials scientists to design microstructures to improve creep resistance and service life prediction of engineering components. However, these theories remain vigorously debated due to a dearth of direct experimental evidence \cite{RN367, RN475, RN436}. Harper-Dorn creep---a mechanism involving intragranular dislocations---and GB sliding have been proposed as alternative mechanisms to explain the creep deformation at low stresses \cite{kumar2007fifty, wadsworth1999deformation}. Therefore, a methodology to predict and measure direct microstructural changes in materials resulting from diffusional creep processes is essential to verify the occurrence of the mechanisms. 

We premise that the diffusion creep process via N-H creep in multicomponent polycrystalline alloys will produce elemental segregation at GBs that is dependent on the GB orientation relative to the applied stress axis. The diffusional creep theories postulated that excess vacancies are produced at GBs normal to the stress axis as they experience tensile stress. In the presence of sufficient thermal activation for vacancy diffusion, the excess vacancies would migrate toward the parallel GBs and get absorbed. We expect such vacancy flux to result in elemental segregation at the GBs when the diffusivities or the diffusive mechanisms of the elements are distinct and the fluxes to the GBs are significant \cite{fuentes1981segregation}. 
Experimental observation of such segregation, validated with theoretical predictions, can provide the necessary evidence for the occurrence of the N-H creep mechanism.
To this end, the present study aims to understand vacancy-mediated diffusion mechanisms and quantify the segregation tendencies of various solutes in Ni-based binary alloys. 

In the last century, Ni-based superalloys have emerged as the preferred candidates for operation in extreme conditions, such as high-temperature creep, fatigue, and highly corrosive environments \cite{pollock_nickel-based_2006, reed_superalloys_2006}. Solutes such as Cr, Re, Ta, W, Mo, and Co contribute to the high-temperature strength of the alloy via solid solution strengthening in the $\gamma$ (matrix) phase and precipitation strengthening via the $\gamma$' (precipitate) phase. Moreover, some of these elements alter the matrix-precipitate interface properties to slow the precipitate coarsening kinetics owing to their diffusion characteristics \cite{giamei_rhenium_1985}. 
Additionally, the kinetic stability of the elements in the phases determines the effectiveness of some of these strengthening mechanisms. Considering these factors, diffusion data and nonequilibrium segregation tendencies are necessary for developing and qualifying superalloys for high-temperature applications. 

Atomistic modeling techniques, such as density functional theory (DFT), have primarily been brought to bear for calculating the energetics of point defect diffusion, such as migration, binding, and formation energies of vacancies. Defect energetics obtained from DFT can be incorporated into higher length scale tools, such as kinetic Monte Carlo (KMC)~\cite{van_der_ven_first_2005} or self-consistent mean field theory (SCMF)\cite{nastar_self-consistent_2000, nastar_mean_2005}. Such tools can compute the total set of Onsager transport coefficients ${L_{ij}}$. 
While there are experimental methods to determine ${L_{ij}}$, these are generally limited to high temperatures and select compositions. Experimental tracer diffusion coefficients have been used to obtain ${L_{ij}}$ using relations such as the Darken or Manning model \cite{manning_diffusion_1961}. 
However, such models cannot predict negative values for off-diagonal components of ${L_{ij}}$ \cite{nastar2007self,messina2014exact}. 
The accurate determination of Onsager transport coefficients is useful not only in determining vacancy and solute diffusivities via the diagonal components but also in the analysis of flux coupling of solute species with vacancies via the off-diagonal components. 
Furthermore, the total set of Onsager transport coefficients is essential to accurately parameterize mean-field and rate-theory models for diffusion creep, thermal nonequilibrium segregation, and radiation-induced segregation ~\cite{senninger_modeling_2016, hargather_comprehensive_2018, messina_solute_2020}.

 There have been several computational studies to calculate solute diffusivities in face-centered cubic (FCC) Ni using DFT energetics.
Solute diffusivities for Cr and Fe were calculated by Tucker et al.~\cite{tucker_ab_2010} and for 26 transition metal solutes by Hargather et al.~\cite{hargather_comprehensive_2018}. These studies \cite{tucker_ab_2010, hargather_comprehensive_2018} employed the five-frequency model developed by Lidiard and LeClaire \cite{lidiard_cxxxiii_1955, leclaire_liii_1956}. The use of analytical methods, such as the five-frequency model in the FCC crystal structure, is computationally efficient and results in reasonable estimates compared to experimental diffusivity measurements~\cite{hargather_comprehensive_2018}. However, this method requires the truncation of solute-vacancy interactions and kinetic correlations beyond the first-nearest neighbor sites. Therefore, it may fail in describing solute-vacancy flux coupling when strong long-range kinetic correlations exist.
One approach to account for these kinetic correlations is the use of the KMC method to calculate the Onsager transport coefficients \cite{van_der_ven_first_2005, arokiam_simulation_2005}. Schuwalow et al. \cite{schuwalow_vacancy_2014} studied the interaction of Re, Ta, W, and Mo solutes with vacancies, and their diffusivities, in FCC Ni by combining the DFT energetics with KMC simulations. This approach is more rigorous compared to the simplified analytical models but is computationally more expensive. As a result, there has always been a motivation to develop a generalized analytical approach to calculate the exact Onsager coefficients with a low computational cost. Such attempts resulted in the development of two methods: the SCMF method by Nastar et al. \cite{nastar_self-consistent_2000, nastar_mean_2005} and the Green's function method by Trinkle \cite{trinkle_automatic_2017}.

Garnier et al.~\cite{garnier_quantitative_2014} used the SCMF method to derive an analytical expression for the Onsager matrix in FCC alloys with solute-vacancy interactions up to the third-nearest neighbor. Their work showed that interactions beyond the first-nearest neighbor sites are important to describe the solute-vacancy flux coupling. Schuler et al.~\cite{schuler_kineclue_2020} implemented the SCMF method in a more generalized way in the KineCluE (kinetic cluster expansion) code by breaking down the transport coefficients into cluster contributions \cite{schuler_transport_2016}. Toijer et al. \cite{toijer_solute-point_2021} utilized DFT calculations and SCMF as implemented in KineCluE~\cite{schuler_kineclue_2020} to study the vacancy and self-interstitial mediated transport behavior of Cr, Fe, Ti, Mn, Si, and P in FCC Ni, with a focus on understanding radiation-induced segregation tendencies. 

A linear elastic approach was developed by Clouet and Varvenne \cite{clouet_dislocation_2008,varvenne_elastic_2017, clouet_elastic_2018} to evaluate the effect of strain on the transport coefficients. This approach was tested by Garnier et al.~\cite{garnier_diffusion_2014} for Si in FCC Ni and by Connetable and Maugis~\cite{connetable_effect_2020} for self-diffusion in FCC Ni. In these studies, the elastic dipole moments were used to calculate the change in migration barriers as a function of strain \cite{garnier_diffusion_2014} or stress \cite{connetable_effect_2020}. Calculations using this approach had an excellent agreement with DFT-based migration barriers calculated on strained supercells \cite{garnier_diffusion_2014, connetable_effect_2020}. The linear elastic approach has also been implemented in the KineCluE code by Schuler et al.~\cite{schuler_kineclue_2020}.

In this work, we employ DFT to obtain the vacancy-mediated diffusion energetics of Cr, Re, Ta, W, Mo, and Co in FCC Ni. By providing the DFT energetics as input into the SCMF framework implemented in the KineCluE code, we evaluate the total set of Onsager transport coefficients, $L_{ij}$, in the limit of dilute solute concentrations.
The Onsager coefficients are then employed to investigate the tendency of solutes for nonequilibrium segregation at vacancy sinks, such as GBs. Furthermore, using the linear elastic approach implemented in KineCluE, we incorporate the effects of tensile hydrostatic strain on the Onsager transport coefficients to obtain further insights into the transport mechanisms and segregation tendencies. These results are then discussed in the context of GBs in polycrystalline materials under diffusion creep conditions. 
In combination with experimental testing in the future, the theoretical predictions from the present work are expected to help verify the hypothesis that nonequilibrium elemental segregation occurs at GBs when the diffusional creep mechanism is active.

\section{Methodology}

This section presents a brief background on the calculation of transport coefficients in \ref{calc_onsager}. The structure of the point defects considered for the calculations and the methodology for calculating the formation, migration, and binding energies of those defects are illustrated in \ref{point_defects}. The DFT computational details are presented in \ref{DFT}.

\subsection{Calculation of Transport Properties}
\label{calc_onsager}

The Onsager transport coefficients ${L_{ij}}$~\cite{allnatt_atomic_2003} relate the flux ($J_i$) of a species $i$ to the chemical potential gradients ($\nabla\mu_j$) of all the species $j$ as follows:
\begin{equation}
   J_i = -\sum_j{L_{ij} \frac{\nabla\mu_j}{k_B T}}. 
\end{equation}
Assuming the vacancy-mediated diffusion mechanisms, the Onsager transport matrix for vacancy and solute components in their dilute limits can be broken down into contributions from isolated vacancies (V) and solute-vacancy pairs (VB) as follows:
\begin{equation}
\label{eq_onsager_matrix}
\begin{bmatrix}
L_{VV} & L_{VB} \\
L_{VB} & L_{BB}
\end{bmatrix}
=
C \Biggl(
f_V 
\begin{bmatrix}
{L_{VV}}^{(V)} & 0 \\
0 & 0 
\end{bmatrix}
+ f_{VB}
\begin{bmatrix}
{L_{VV}}^{(VB)} & {L_{VB}}^{(VB)} \\
{L_{VB}}^{(VB)} & {L_{BB}}^{(VB)}
\end{bmatrix}
\Biggr),
\end{equation}
where the superscript (V) represents that the transport coefficient is calculated from a cluster with a monovacancy and the superscript (VB) represents that the coefficient is calculated from a cluster with a solute-vacancy pair. $C$ is the total concentration
of defect clusters (monovacancies and solute-vacancy pairs in this work). 
$f_V$ and $f_{VB}$ are the cluster fractions of monovacancies and solute-vacancy pairs, respectively. Each of these fractions is proportional to the respective partition function calculated by KineCluE using the SCMF theory \cite{nastar_self-consistent_2000, nastar_mean_2005, schuler_kineclue_2020}. 

With $L_{VV}$, $L_{VB}$, and $L_{BB}$ obtained, we can use the Onsager reciprocity relations $L_{ij}=L_{ji}$ and the conservation of substitutional lattice sites via $J_A+J_B+J_V=0$ to compute the transport coefficients related to the solvent atom (A) as follows \cite{van2010vacancy}:
\begin{equation}
\label{eq_LAB}
    L_{AB} = -L_{VB} - L_{BB},
\end{equation}
\begin{equation}
\label{eq_LAA}
    L_{AA} = L_{VV} - 2L_{AB} - L_{BB},
\end{equation}
\begin{equation}
\label{eq_LAV}
    L_{AV} = -L_{AA} - L_{AB}.
\end{equation}

The evaluation of cluster fractions is discussed briefly in Appendix \ref{appendix_partition}. A thorough discussion on the evaluation of transport coefficients and partition functions using KineCluE has been provided by Schuler et al.~\cite{schuler_kineclue_2020} and Messina et al.~\cite{messina_solute_2020}. 

\subsection{Structure and Energetics of Point Defects}
\label{point_defects}
\begin{figure}[htp!]
    \centering
    \includegraphics[scale=0.7]{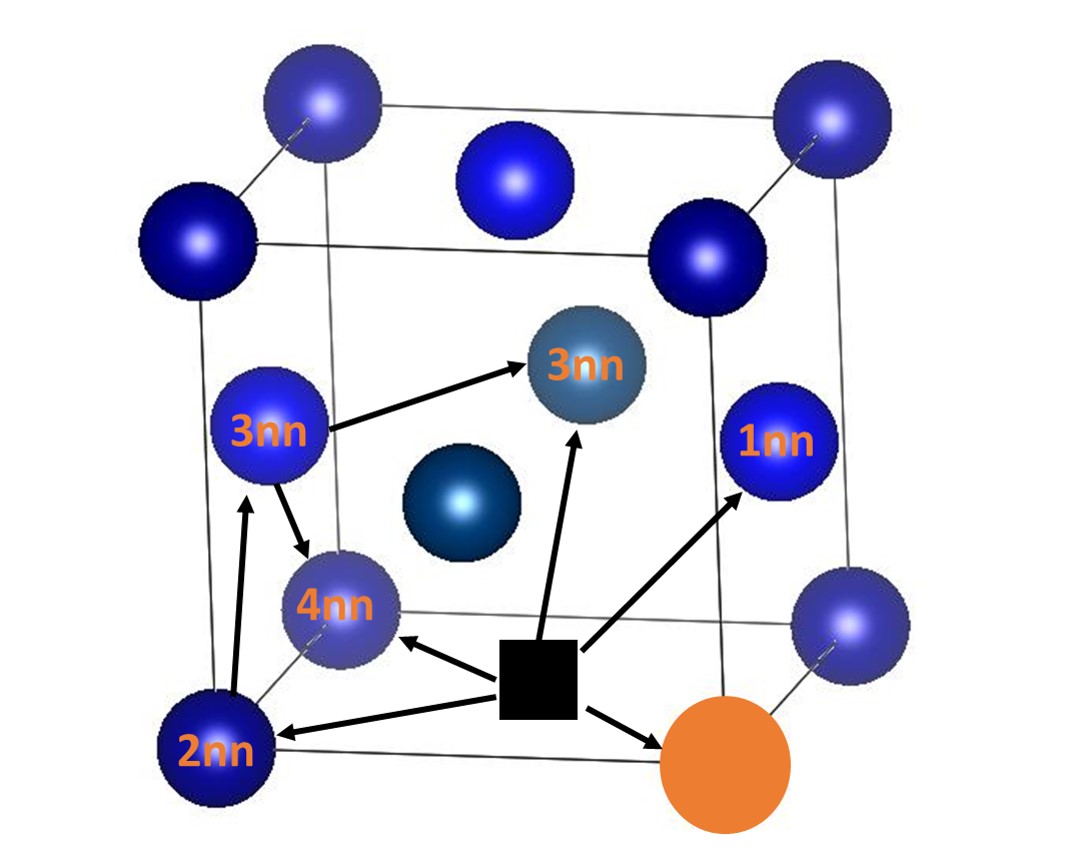}
    \caption{The vacancy jumps considered in the FCC Ni system. The blue spheres are nickel atoms. The solute atom is the orange sphere at the corner, and the black square is a vacancy at the first-nearest neighbor (nn) distance from the solute atom. The arrows correspond to jumps from one lattice site to another when one of them is vacant. }
    \label{fig:fcc_structure}
\end{figure}

Diffusion via vacancy migration is first considered to model the self and solute diffusion in FCC Ni. The calculation of cluster transport coefficients for monovacancies and solute-vacancy pairs requires the solute-vacancy binding energies and migration energies for all possible jumps within a defined thermodynamic range. This thermodynamic range is taken to be $\sqrt{2}a$ from a solute atom, where $a$ is the lattice parameter of FCC Ni. Any solute-vacancy interaction separated by a distance more than $\sqrt{2}a$ will be neglected (i.e., zero binding energy). Similarly, any vacancy jump to a solvent atom occurring beyond that range from a solute atom will be treated as a monovacancy jump. The choice of this value for the thermodynamic range is attributed to the convergence of vacancy-binding energies to almost zero at a distance beyond the fourth-nearest neighbor distance, as will be shown in Section \ref{results}. This thermodynamic range is distinct from the kinetic range defined in KineCluE code \cite{schuler_kineclue_2020}. The kinetic range defines the maximum extent of the kinetic trajectories included in the SCMF calculation. In this work, we adopted a value of $4a$ for the kinetic range after performing convergence tests with values of $2a$, $3a$, $4a$, and $5a$.
The solute-vacancy configurations and the possible jumps in the FCC structure considered in this work are shown in Figure \ref{fig:fcc_structure}.

This subsection presents the governing equations used to calculate the point defect energetics. First, the vacancy formation energy $E_{f}^{V}$ is calculated as follows:
\begin{equation}
\label{eq_E_F_vac}
E_{f}^{V} = E^{V}_{defect}[(N-1)Ni] - \frac{N-1}{N} E_{perfect}[(N) Ni],
\end{equation}
where $E^{V}_{defect}[(N-1)Ni]$ and  $E_{perfect}[(N) Ni]$  are the DFT energies of the defective and perfect supercells, respectively. The formation energies of a solute substitutional $E_{f}^{sub,B}$ and of a solute-vacancy pair $E_{f}^{VB}$ are given by:
\begin{equation}
\label{eq_E_F_sub}
E_{f}^{sub,B} = E^{sub,B}_{defect}[(N-1)Ni + B] - \frac{N-1}{N} E_{perfect}[(N) Ni] - E_{B},
\end{equation}
\begin{equation}
\label{eq_E_F_VB}
E_{f}^{VB} = E^{VB}_{defect}[(N-2)Ni + B] - \frac{N-2}{N} E_{perfect}[(N) Ni] - E_{B},
\end{equation}
where $E^{sub,B}_{defect}[(N-1)Ni + B]$ and $E^{VB}_{defect}[(N-2)Ni + B]$ are the DFT energies of supercells containing an isolated substitutional and a solute-vacancy pair, respectively. $E_B$ is the DFT energy per atom of element $B$ in its pure crystalline form.
The solute-vacancy binding energies ($E_{b}^{VB}$) are calculated from the formation energies in Equations \ref{eq_E_F_vac}, \ref{eq_E_F_sub}, and \ref{eq_E_F_VB} according to the following formula:
\begin{equation}
E_{b}^{VB} = E_{f}^{V} + E_{f}^{sub,B} - E_{f}^{VB}.
\end{equation}
According to this formula, positive binding energy means attraction and negative binding energy means repulsion. 

The jump frequency ($\omega^{i\rightarrow f}$) from an initial configuration (i) to a final configuration (f) is given by: 
\begin{equation}
    \omega^{i\rightarrow f} = \nu^{i\rightarrow f} e^{\frac{-E_m^{i\rightarrow f}}{k_B T}},
\end{equation}
where $E_m^{i\rightarrow f}$ is the migration energy for the jump from $i$ to $f$. The migration energies are calculated from the differences between the ground state energy of the initial configuration and the saddle point energy on the minimum energy path between $i$ and $f$. $\nu^{i\rightarrow f}$ is the attempt frequency and is evaluated from the phonon frequencies of the initial relaxed state and the transition state (saddle point) of the jump according to Vineyard's harmonic transition state theory \cite{vineyard_frequency_1957}. We use an approximate form of the full Vineyard expression as follows:
\begin{equation}
    \nu^{i\rightarrow f} = \frac{\Pi^{3N_{hop}}_j  \nu^{i}_j}{\Pi^{3N_{hop}-1}_j  \nu^{s,i\rightarrow f}_j}.
\end{equation}
According to this equation, we only consider the hopping Ni or solute atom (i.e., $N_{hop} = 1$) instead of considering the phonon frequencies of every atom in the supercell, which would be computationally very expensive. Finally, we calculate the vacancy formation entropy $S_{f}^{V}$ from the phonon frequencies of the perfect and defective (with a vacancy) supercells, $\nu_j^{perfect}$ and $\nu_j^{defect}$, respectively, as follows:
\begin{equation}
    S_{f}^{V} = k_B ln(\frac{\Pi^{3N-1}_j \nu_j^{perfect}}{\Pi^{3N-1}_j \nu_j^{defect}}).
\end{equation}
We also reduce the computational expense of this phonon calculation by only considering the vibrational frequencies of the first-nearest neighbor atoms to the vacancy (i.e., 12 neighboring atoms).

The binding energies and jump frequencies calculated for different solute-vacancy configurations (See Figure \ref{fig:fcc_structure}) in this section constitute the required input data to the KineCluE code \cite{schuler_kineclue_2020} for calculating the Onsager transport coefficients. 

\FloatBarrier

\subsection{Density Functional Theory Computational Details}
\label{DFT}
DFT calculations are performed using the projector-augmented wave method \cite{blochl_projector_1994} as implemented in the Vienna ab initio Simulation Package \cite{kresse_ab_1993, kresse_efficient_1996}. Spin-polarized calculations with collinear magnetic moments with an initial value of 2 Bohr magnetons ($\mu$B) per atom were implemented to generate a ferromagnetic Ni system, The relaxed FCC Ni structure has a magnetic moment of 0.6  $\mu$B per atom. The same value of 2 $\mu$B  was used to initialize the magnetic moment of the solute atoms, Co, Re, Mo, Ta, and W which yielded a final magnetic moment value of 1.8, -0.1, -0.2, -0.2, -0.2  $\mu$B, respectively. The negative sign denotes a spin-down state (opposite to that of the Ni atoms). In the case of Cr solute, initial guesses of spin-up (2  $\mu$B ) and spin-down (--2 $\mu$B ) states were tested but both yielded the same final magnetic moment of 2.1  $\mu$B.
The Perdew-Burke-Ernzerhof \cite{perdew_generalized_1996} generalized-gradient approximation (GGA) is adopted for the exchange-correlation functional. A plane-wave cutoff energy of 450 eV was set in all calculations. The convergence criteria for total energies and forces were set to $10^{-6}$ eV and 0.005 eV/\AA, respectively, for relaxing defect-containing supercells at constant volume. For climbing-image nudged elastic band {(CI-NEB)} calculations \cite{mills_quantum_1994, mills_reversible_1995,henkelman_climbing_2000}, one intermediate image was used to obtain the saddle point energies along migration paths.
{Typically, a single intermediate image is sufficient for a vacancy jump; this provides fast convergence of the saddle point energies and is commonly used in the literature }{\cite{agarwal_exact_2017, jain_first-principles_2019}.}{ Additionally, three-image CI-NEB calculations were also tested for the Cr and Re solute-vacancy exchange jumps. The obtained saddle points agree with the single-image calculations to within 1 meV.}
The convergence criteria in {CI-NEB} calculations were set to $10^{-5}$ eV and 0.01 eV/\AA, for total energies and forces, respectively. $3\times3\times3$ (108 atoms) and $4\times4\times3$ (192 atoms) supercells were used for calculating point defect formation, binding, and migration energies. $5\times5\times5$ and $4\times4\times5$ k-point meshes were used for the Brillouin zone sampling of the 108-atom and 192-atom supercells, respectively. The results from both supercell sizes agree well for all solute species, except for Cr-vacancy pairs where the 108-atom supercell gives overestimated migration barriers for certain vacancy jumps (especially the $1nn\rightarrow1nn$ jump). 
The migration barriers obtained for the Ni-Cr system from the 192-atom supercell have better agreement with the results reported by Toijer et al.~\cite{toijer_solute-point_2021} using 256-atom supercells. This gives us confidence in our 192-atom results. The results presented in this work are based on the 192-atom supercell DFT data unless noted otherwise.

The phonon calculations required for the evaluation of attempt frequencies and vacancy formation entropy were performed by calculating the force constants using the finite-differences approach as implemented in the Vienna ab initio Simulation Package~\cite{kresse_ab_1993,kresse_efficient_1996}. Atoms were displaced by 0.02 {\AA} to calculate the Hessian matrix~\cite{hafner_ab-initio_2008}.

\section{Results and Discussion}
\label{results}

\subsection{Density Functional Theory Results}
\label{DFT_results}

The bulk properties calculated in pure FCC Ni are tabulated in Table \ref{tab:bulk_prop}. The vacancy formation energy and entropy are 1.42 eV and 1.89 $k_B$, respectively. Both are in good agreement with previously calculated values in the literature \cite{schuwalow_vacancy_2014,toijer_solute-point_2021,yu_effect_2009,tucker_ab_2010}. It is known that GGA may underestimate vacancy formation energies in metals \cite{medasani_vacancy_2015}. An approach suggested by Tucker et al.~\cite{tucker_ab_2010} is to use the experimental vacancy formation energy (1.79 eV \cite{ehrhart1991atomic}) to reproduce the vacancy diffusion coefficients. This approach is not followed in this work since the main goal is not to match experimental diffusion coefficients but to study the solute segregation tendencies which are independent of the vacancy formation energy.
The vacancy migration energy of 1.09 eV is also in excellent agreement with previous DFT calculations of 1.01 \cite{schuwalow_vacancy_2014}, 1.09 \cite{tucker_ab_2010}, and 1.05 \cite{toijer_solute-point_2021} eV.

\begin{table}[htp!]
    \caption{Bulk properties of FCC Ni calculated in this work, compared with experiments~\cite{hermann_appendix_2011} and previous atomistic calculations~\cite{schuwalow_vacancy_2014, toijer_solute-point_2021, yu_effect_2009, goswami_can_2014, tucker_ab_2010}.}
    \centering
    \begin{tabular}{|c|c|c|}
    \hline
     Quantity & This Work & Literature \\
    \hline
    Lattice constant (\AA)& 3.516 & 3.52 \cite{hermann_appendix_2011}, 3.528 \cite{schuwalow_vacancy_2014}, 3.522 \cite{toijer_solute-point_2021}  \\
    Vacancy formation energy (eV)& 1.42 &1.42 \cite{schuwalow_vacancy_2014}, 1.40 \cite{toijer_solute-point_2021}, 1.39 \cite{yu_effect_2009} \\
    Vacancy formation entropy ($k_B$)& 1.82 & 1.89 \cite{tucker_ab_2010} \\
    Vacancy migration energy (eV) & 1.09 & 1.01 \cite{schuwalow_vacancy_2014}, 1.09 \cite{tucker_ab_2010}, 1.05 \cite{toijer_solute-point_2021} \\
    Attempt frequency (THz) & 4.65 & 2.58 \cite{goswami_can_2014}, 4.48 \cite{tucker_ab_2010}, 4.63 \cite{ke2019ab}, 4.8 \cite{garnier_diffusion_2014} ,14.32 \cite{toijer_solute-point_2021} \\
    \hline
    \end{tabular}
    \label{tab:bulk_prop}
\end{table}

\begin{figure}
    \centering
    \includegraphics[scale=0.8]{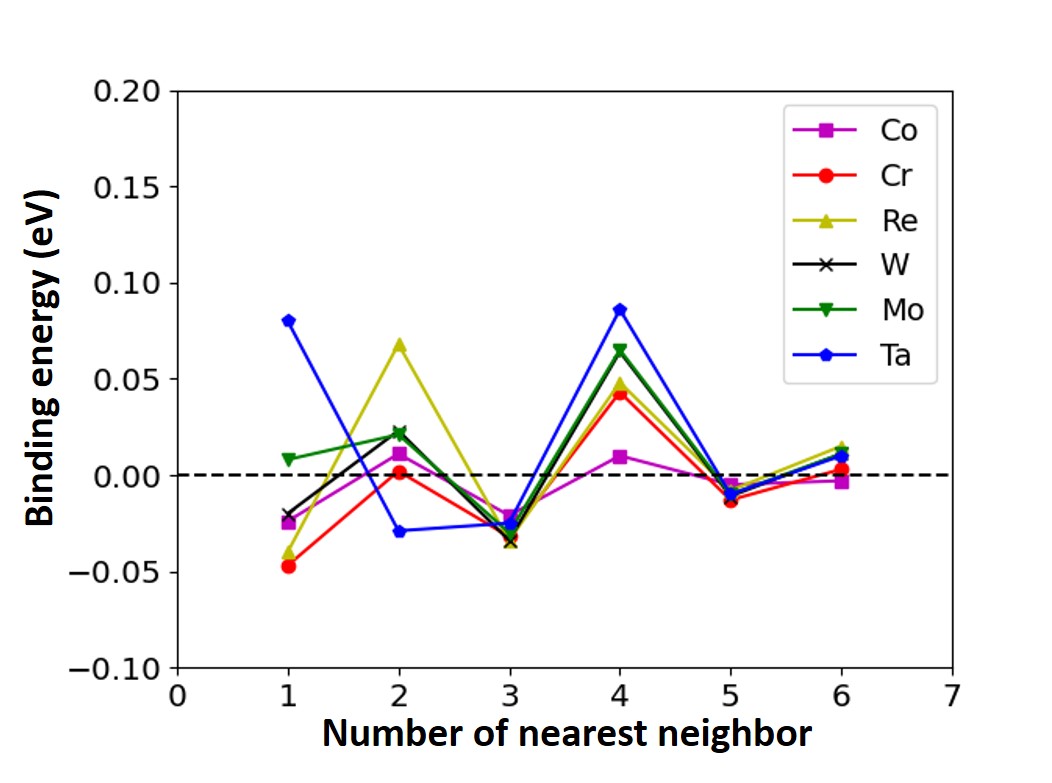}
    \caption{Solute-vacancy binding energies as a function of nearest neighbor distance in FCC nickel. {The nearest neighbor number refers to the position of vacancy with respect to the solute atom.} Positive binding energies are attractive and negative binding energies are repulsive.}
    \label{fig:binding_energies}
\end{figure}

The solute-vacancy binding energies are plotted in Figure \ref{fig:binding_energies} for separation distances ranging from the first- to the sixth-nearest neighbor. The numerical values for these binding energies are listed in Table \ref{tab:binding_energies} in Appendix \ref{appendix_DFT_data}. The magnitude of the binding energies is relatively low (within $\pm$0.05 eV) for all the solutes considered in this work except for Ta. For Cr, Re, W, and Co, the solute-vacancy interaction at the first-nearest neighbor distance is repulsive while it is attractive at the second-nearest neighbor distance. The same trends have been previously observed for Re and W in FCC Ni by Schuwalow et al. \cite{schuwalow_vacancy_2014} and for Cr in FCC Ni by Toijer et al. \cite{toijer_solute-point_2021}. Ta shows the opposite of this behavior where it exhibits an attractive binding with vacancy at the first-nearest neighbor and a repulsive interaction at the second-nearest neighbor distance. On the other hand, Mo has a weak attraction with the vacancy at both first- and second-nearest neighbor distances. The differences in the solute-vacancy binding energies for different solutes can be attributed to the nature of bonding between solute and solvent atoms and how vacancies affect that bonding. The electronic origin of such differences in binding energies and migration energies has been studied before in FCC Ni \cite{janotti2004solute, ke2019ab, ke2022first} and in body-centered cubic Fe~\cite{ohnuma2009first, messina_systematic_2016} alloys, but is beyond the scope of the present work. 

\begin{figure}
    \centering
    \includegraphics[scale=0.8]{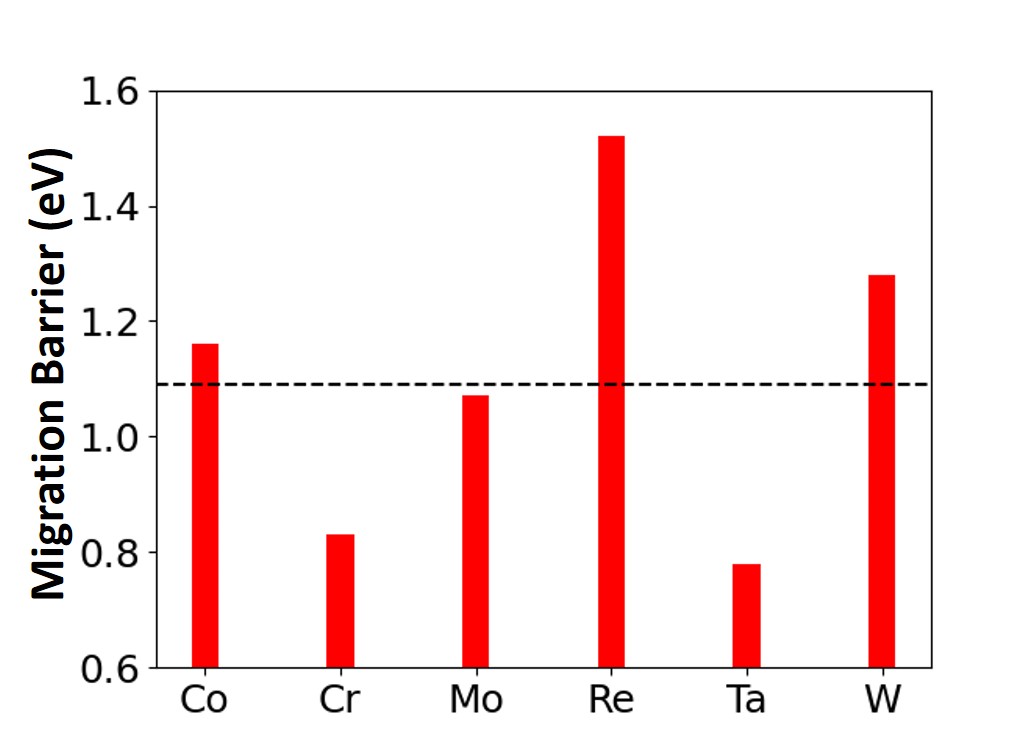}
    \caption{The calculated migration barriers of solute-vacancy exchange jumps in FCC Ni. The dashed black line indicates the solute-vacancy exchange barrier (self-diffusion barrier) of 1.09 eV for comparison.}
    \label{fig:migration_barriers}
\end{figure}

The migration energies for the solute-vacancy exchange jumps are shown in Figure \ref{fig:migration_barriers}. The full list of migration energies for all the jumps shown in Figure \ref{fig:fcc_structure} are listed in Table \ref{tab:migration_energies} in Appendix \ref{appendix_DFT_data} for each of the six solutes. In Figure \ref{fig:migration_barriers}, the solute migration energies are compared against the self-migration energy (for solvent-vacancy exchange), which is 1.09 eV. 
Cr, Ta, and Mo have relatively low migration barriers of 0.83, 0.78, and 1.07 eV, respectively, for the solute-vacancy exchange.
Due to the lower migration energies compared to the self-migration energy of Ni, we expect Cr, Ta, and Mo to be fast diffusers in FCC Ni. 
On the other hand, Re, W, and Co have relatively high migration barriers of 1.52, 1.28, and 1.16 eV, respectively. 
Due to this, we expect Re, W, and Co to be slow diffusers.
We note that, in addition to the solute-vacancy migration barriers, the attempt frequencies and solute-vacancy binding energies need to be taken into consideration to conclusively assess the fast or slow nature of solute mobilities at a given temperature. 
While the full picture will be presented in Section \ref{diffusivites_and_drag_ratios}, the above classification of Cr, Ta, and Mo as fast diffusers and Re, W, and Co as slow diffusers doesn't change.

Table \ref{tab:attempt_freq} presents the attempt frequencies for solute-vacancy exchange jumps. The attempt frequencies for other jumps in Table \ref{tab:migration_energies} (association, dissociation, or reorientation jumps) will be assumed to be the same as the attempt frequency of Ni self-diffusion (i.e., 4.65 THz) to reduce computational expense. This is a reasonable approximation because the solute displacement in these jumps is negligible and the residual strain caused by the solute is expected to cause a minimal variation in the phonon frequencies of the solvent Ni atoms. The attempt frequencies calculated in this work show a good agreement with literature values reported by Tucker et al. \cite{tucker_ab_2010} for both Ni self-diffusion and Cr-vacancy exchange as shown in Tables \ref{tab:bulk_prop} and \ref{tab:attempt_freq}, respectively. The values reported by Toijer et al. \cite{toijer_solute-point_2021} and Goswami et al. \cite{goswami_can_2014} are significantly higher and lower, respectively. This scatter in the attempt frequency data in the literature calls for a more accurate investigation of the methods used to calculate attempt frequencies using DFT. However, this is beyond the scope of this work and such differences in attempt frequencies are not expected to result in large disagreements in transport coefficients. The transport coefficients are more sensitive to changes in migration energies than to changes in prefactors.

\begin{table}[htp!]
    \caption{Attempt frequencies in THz for solute-vacancy exchange in FCC nickel obtained from the 108-atom supercells.}
    \centering
    \begin{tabular}{|c|c|c|c|c|c|c|c|}
    \hline
         & Cr &  Re & Ta & W & Mo & Co \\
    \hline
       This work &4.81 &3.27 &3.03 &3.21 &4.06 &4.59\\
    \hline
    Literature (DFT) &4.92 \cite{tucker_ab_2010}, 10.85\cite{toijer_solute-point_2021} & 1.67 \cite{goswami_can_2014}&2.56 \cite{goswami_can_2014} &2.36 \cite{goswami_can_2014} &--- &---\\
    \hline
    \end{tabular}
    \label{tab:attempt_freq}
\end{table}

\subsection{Solute Diffusion Coefficients}
\label{diffusivites_and_drag_ratios}

\begin{figure}[htp!]
    \centering
    \includegraphics[scale=0.48]{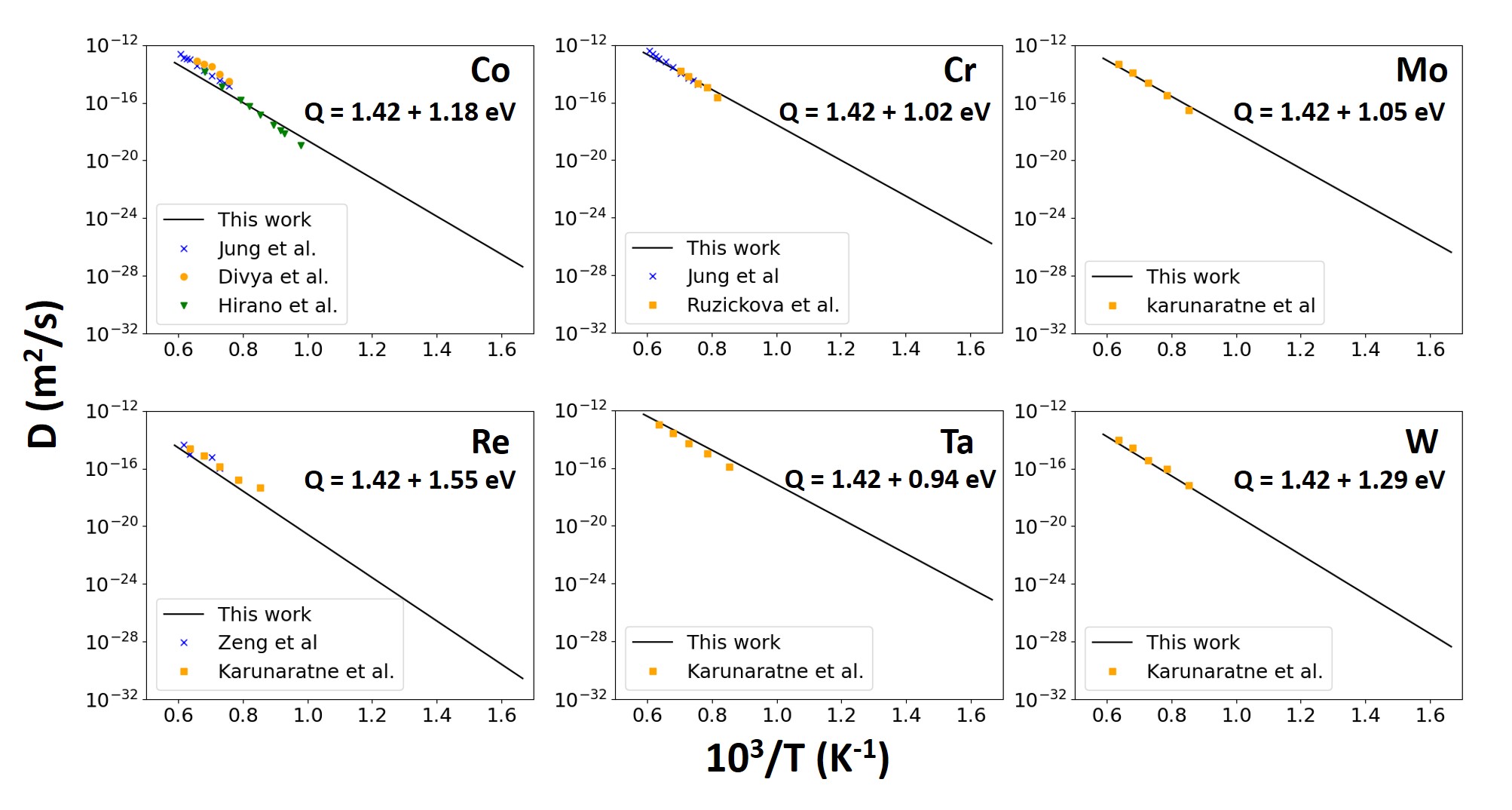}
    \caption{Vacancy-mediated diffusion coefficients of the solutes Co, Cr, Mo, Re, Ta, and W in FCC Ni are plotted versus inverse temperature. Activation barriers (Q) are shown as a sum of two terms, where the first term is the contribution from the vacancy formation energy (1.42 eV) and the second term includes the contributions from binding energies and migration barriers. Calculations from this work are plotted in solid lines; experimental diffusivities from Refs. \cite{jung_interdiffusion_1992, divya_diffusion_2011, hirano_diffusion_1962, ruzickova_self-diffusion_1981, karunaratne_interdiffusion_2000, karunaratne_interdiffusion_2005, zeng_study_2009} are shown by the scattered markers.}
    \label{fig:solute_diffusivities}
\end{figure}

Starting from the DFT data reported in Section \ref{DFT_results}, the SCMF approach as implemented in KineCluE~\cite{schuler_kineclue_2020} is used to obtain the transport coefficients as explained in Section \ref{calc_onsager}. The diffusion coefficient ($D_B$) of a solute B is calculated from the $L_{BB}$ component of the transport coefficient matrix and the solute concentration $[B]$ as:
\begin{equation}
    D_B = \frac{L_{BB}}{[B]}.
\end{equation}
The solute diffusivities calculated using this approach are valid for dilute concentrations of $[B] < 0.094\%$ (see Appendix \ref{appendix_partition}). The diffusivities of the six solutes are plotted versus inverse temperature in Figure \ref{fig:solute_diffusivities}. Experimental tracer diffusivity data from Refs.~\cite{jung_interdiffusion_1992, divya_diffusion_2011, hirano_diffusion_1962, ruzickova_self-diffusion_1981, karunaratne_interdiffusion_2000, karunaratne_interdiffusion_2005, zeng_study_2009} are plotted alongside for comparison. The results show a reasonable agreement between the KineCluE calculations and the experimental data. The calculated diffusivities were fit to the Arrhenius equation $D = D_0 e^{-Q/{k_B T}}$ to determine the activation energies $Q$ and the prefactors $D_0$, which are listed in Table \ref{tab:arrhenius_fit}. We also performed these calculations for the solvent Ni.
The activation energy for Ni self-diffusion was obtained as 2.51 eV (i.e., $Q = E_f^V + E_{m}^V$  = 1.42 eV + 1.09 eV).
Amongst the solutes, Re, W, and Co (the slow diffusers) yielded higher effective activation energies of 2.97, 2.71, and 2.60 eV, respectively, while Mo, Cr, and Ta (the fast diffusers) yielded low activation energies of 2.47, 2.44, and 2.36 eV, respectively.{ Comparing the values of $Q$ with the migration barriers of solute-vacancy exchange ($E_{m}^{V-B}$) in Table\mbox{~\ref{tab:migration_energies}} and Figure\mbox{~\ref{fig:migration_barriers}}, we find a strong correlation. This is because Q can be roughly estimated by adding $E_f^V$---which remains constant regardless of the solute---and $E_{m}^{V-B} - E_b^{1nn}$ under the assumption of ideal random walk behavior\mbox{~\cite{LECLAIRE197870}}; here, $E_b^{1nn}$ is the binding energy of the solute B with a vacancy in the first nearest neighbor position. The above indicates that the solute-vacancy exchange controls the vacancy diffusion mechanism of these solutes and that deviations from random walk behavior due to correlation effects are minimal, especially for Co, Mo, Re, and W. For the two fastest diffusers, Cr and Ta, kinetic correlation effects are more prevalent and the sum of the formation, binding, and migration energies does not accurately equal the exact value of Q; this difference is the activation energy of the correlation factor.}

\begin{table}[htp!]
    \centering
    \caption{Activation energies (Q) and diffusion prefactors ($D_0$) obtained in this work are tabulated and compared against previous theoretical and experimental calculations in the literature.}
    \setlength\extrarowheight{-3pt}
    \label{tab:arrhenius_fit}
    \begin{tabular}{|c|c|c|c|c|}
    \hline
       Solute & Method & Q (eV) & Q (kJ/mol)  & $D_0$ ($\times 10^{-6} m^2/s$ )  \\
       \hline
        Co &GGA + SCMF (this work) &2.60 & 251 &3.10 \\
         & LDA + 5-frequency \cite{hargather_comprehensive_2018} &3.07--3.09 & 296--298 &8.5--24.3 \\
         & Exp. \cite{jung_interdiffusion_1992} &2.92 &282 & 180 \\
         & Exp. \cite{divya_diffusion_2011} &2.84 &274 &  ---  \\
         & Exp. \cite{hirano_diffusion_1962} &2.81 & 271 & 75\\
         \hline
         Cr & GGA + SCMF (this work) &  2.44 & 235 &5.36 \\
         & GGA + 5-frequency \cite{tucker_ab_2010} & 2.64 & 255 & 4.52  \\
         & LDA + 5-frequency \cite{hargather_comprehensive_2018} &3.06--3.11 & 295--300 & 15.1--92.5\\ 
         & Exp. \cite{jung_interdiffusion_1992} & 3.00 & 289 & 520 \\
         & Exp. \cite{ruzickova_self-diffusion_1981} &3.03  & 292 & 93 \\
         \hline
         Mo & GGA + SCMF (this work) &2.47 &239 & 2.56\\
         & GGA + KMC \cite{schuwalow_vacancy_2014} &2.44 &235 &0.96 \\
         & LDA + 5-frequency \cite{hargather_comprehensive_2018} & 2.92--2.90 &282--280 & 4.9--10.9 \\
         & Exp. \cite{karunaratne_interdiffusion_2005} & 2.92 &281 & 115 \\
         \hline
         Re & GGA + SCMF (this work) &2.97 &287 & 2.65\\
         & GGA + KMC \cite{schuwalow_vacancy_2014} &2.91 &281 &1.04 \\
         & LDA + 5-frequency \cite{hargather_comprehensive_2018} & 3.57--3.54&345--342 &4.6--8.8 \\
         & Exp. \cite{zeng_study_2009} &4.27 &412 & --- \\
         & Exp. \cite{karunaratne_interdiffusion_2000} &2.64 &255 &0.82 \\
         \hline
         Ta & SCMF (this work) &2.36 &228 & 5.38\\
         & GGA + KMC \cite{schuwalow_vacancy_2014} &2.28 &220 &1.54 \\
         & LDA + 5-frequency \cite{hargather_comprehensive_2018} & 2.82--2.66&272--257 &18.5--39.7 \\
         & Exp. \cite{karunaratne_interdiffusion_2000} &2.60 &251 &21.9 \\
         \hline
         W & GGA + SCMF (this work) &2.71 & 261 &2.46 \\
         & GGA + KMC \cite{schuwalow_vacancy_2014} &2.66 &257 &1.10 \\
         & LDA + 5-frequency \cite{hargather_comprehensive_2018} & 3.13--3.09&302--299 & 3.7--6.9 \\
         & Exp. \cite{karunaratne_interdiffusion_2000} &2.74 &264 & 8.0 \\
         \hline
    \end{tabular}
\end{table}

\subsection{Solute-vacancy Drag and Segregation Tendencies} \label{sec:drag_segregation}

The solute-vacancy flux couplings and solute segregation tendencies can be analyzed from the vacancy drag and partial diffusion coefficient ratios. 
The vacancy drag ratio $G_V$ is a measure of whether the solute diffuses in the same direction as the vacancies ($G_V > 0$) or in the opposite direction ($G_V < 0$) also referred to as the inverse Kirkendall effect (IKE) \cite{marwick_segregation_1978}. The vacancy drag ratio $G_V$ is defined as follows:
\begin{equation}
    G_V = \frac{L_{VB}^{(VB)}}{L_{BB}^{(VB)}}.
\end{equation}
The partial diffusion coefficient ratio $D_{pd}$ is a measure of the solute diffusion relative to that of the solvent Ni atoms and is defined (for vacancies) as follows \cite{nastar20121, wolfer1983drift}:
\begin{equation}
    D_{pd} = \frac{(1-[B])}{[B]} \frac{L_{VB}}{L_{AV}}.
\end{equation}
While partial fluxes and diffusion coefficients were originally introduced to distinguish contributions from vacancies and self-interstitial atoms, we only consider the contributions from vacancies since self-interstitial concentrations are expected to be negligible under thermal conditions.
Under ideal conditions---where other sinks or sources for vacancies and trapping or clustering sites for vacancies and solutes are absent---the value of $D_{pd}$ indicates whether enrichment or depletion of the solute occurs at vacancy sinks. 

\begin{figure}[htp!]
    \centering
    \includegraphics[scale=0.13]{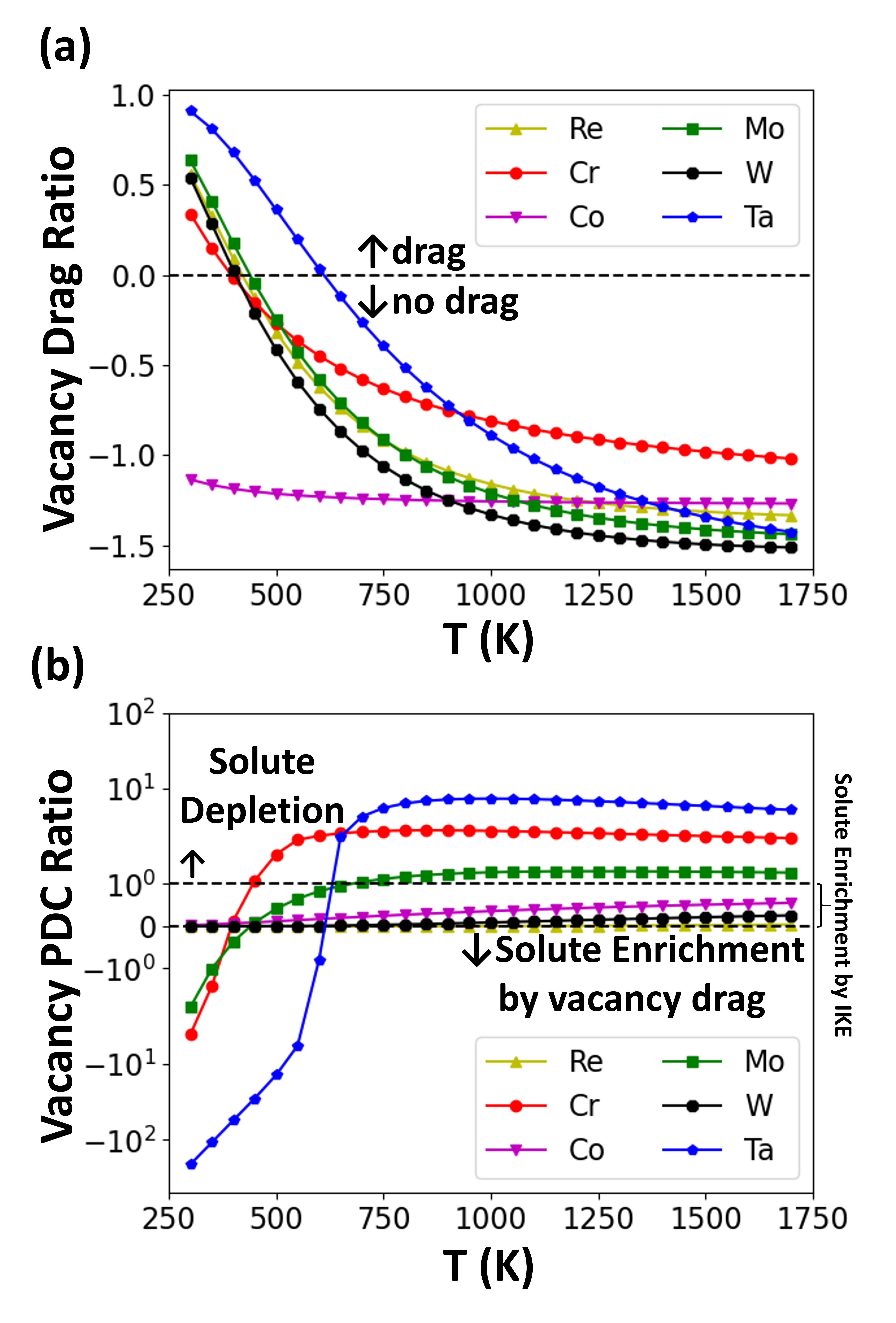}
    \caption{(a) Vacancy drag ratios, $G_V$ and (b) partial diffusion coefficient (PDC) ratios, $D_{pd}$, for solutes in FCC Ni on a semilogarithmic plot. PDC ratios $>$1 indicate solute depletion at sinks, while PDC ratios $<$0 indicate solute enrichment via the vacancy drag mechanism. PDC ratios between 0 and 1 also indicate solute enrichment but via the solute-vacancy exchange mechanism and IKE.}
    \label{fig:drag}
\end{figure}

$G_V$ and $D_{pd}$ are plotted versus temperature in Figures \ref{fig:drag}a and \ref{fig:drag}b, respectively. $D_{pd}$ was evaluated at the maximum solute concentration of $[B] = 0.094\%$ allowed in the present dilute-limit model (See Appendix \ref{appendix_partition}).
Figure \ref{fig:drag}a shows that all the considered solutes in FCC Ni have negative $G_V$ at high temperatures. That is, the solutes move away from vacancy sinks, in a direction opposite to that of the vacancies. With decreasing temperature, $G_V$ keeps increasing and transitions to positive values at a temperature, $T_{drag}$, where $G_V = 0$. Below $T_{drag}$, the solutes are dragged by vacancies. The drag temperatures $T_{drag}$ for all the solutes are reported in Table \ref{tab:drag_transition}.
Note that Co is dragged by vacancies only below 20 K. 

Although the vacancy drag behavior of the six solutes is very similar (especially at high temperatures above 700 K), their segregation tendencies can be very different, as inferred from $D_{pd}$ in Figure~\ref{fig:drag}b. 
This is because the segregation tendencies for $G_V < 0$ can either be enrichment or depletion, depending on the transport of the solute relative to Ni. 
With decreasing temperature, the transition from depletion to enrichment occurs at a crossover temperature $T_e$.
When $G_V > 0$, on the other hand, the solute always enriches since it is dragged by the vacancies, whereas Ni is transported in an opposite direction via vacancy exchange.
Therefore, for solutes (fast diffusers) that undergo a crossover in segregation, we observe $T_e > T_{drag}$ (Table \ref{tab:drag_transition}).

\begin{figure}[htp!]
    \centering
    \includegraphics[scale=0.7]{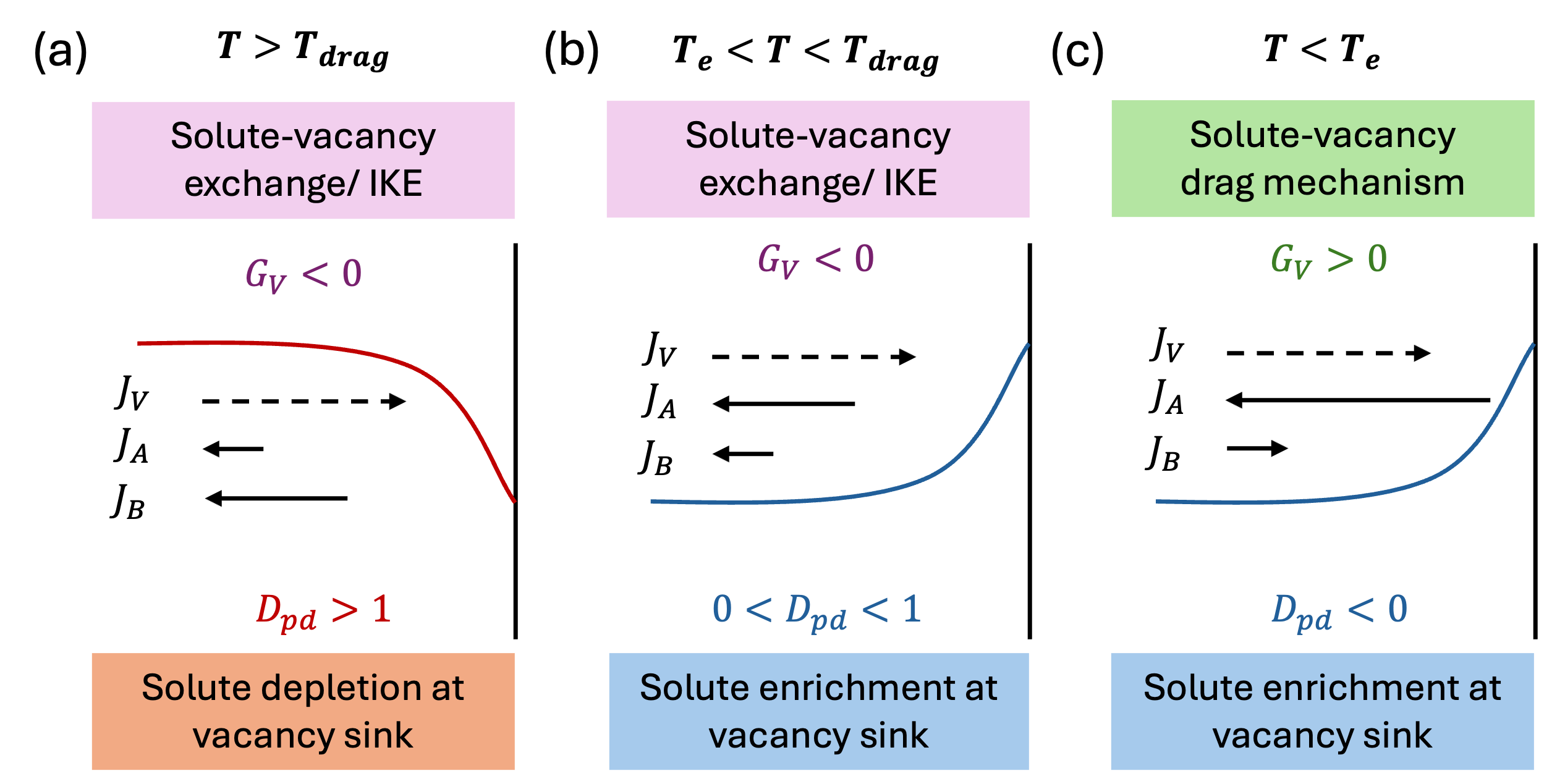}
    \caption{Schematic of solute (B) segregation profiles near a GB (vertical line) acting as a vacancy sink. Arrows indicate the direction of fluxes of various components. Conditions and mechanisms under which solute depletion or enrichment occur are summarized.}
    \label{fig:schematic}
\end{figure}

From Figure~\ref{fig:drag}b, we infer that the fast diffusers (Ta, Cr, and Mo) tend to deplete ($D_{pd} > 1$) at high temperatures above $T_e$; this is due to the more favorable solute-vacancy exchange compared to the solvent-vacancy exchange. 
At lower temperatures, in the range where $0 < D_{pd} < 1$ and $T_{drag} < T < T_e$, the net solute flux (Ta, Cr, and Mo) is still in the opposite direction of the vacancy flux.
However, the net solute flux away from the sinks is lower than that of Ni atoms flux ($0 < \frac{L_{VB}}{L_{AV}} < 1$) as illustrated in Figure~\ref{fig:schematic}b.
Although not dominant, vacancy drag likely contributes to reducing the net solute flux, resulting in an enrichment tendency of the solute at sinks.
We also note that the fast diffusers exhibit a narrow window of temperature between $T_{drag}$ and $T_e $ where $0<D_{pd}<1$.
The vacancy drag ratios increase with decreasing temperature, and therefore,
at even lower temperatures (i.e., below $T_{drag}$), the vacancy drag mechanism dominates as illustrated in Figure~\ref{fig:schematic}c, resulting in $G_V > 0$ and $D_{pd} < 0$. At these temperatures, Ta, Cr, and Mo are enriched at sinks due to vacancy drag.

\FloatBarrier

In contrast, the slow diffusers (Re, W, and Co) always exhibit $D_{pd} < 1$. 
At high temperatures, when $G_V < 0$, the solutes diffuse slowly compared to the solvent atoms. 
While they tend to move away from the sinks (via solute-vacancy exchange or IKE), they have significantly slower diffusivity than Ni. Consequently, since the Ni-vacancy exchange is more favorable than the solute-vacancy exchange ($0 < D_{pd} < 1$), the solutes are left behind and effectively enriched at vacancy sinks \cite{marwick_segregation_1978}. 
There is no calculated crossover temperature $T_e$ for these solutes because $D_{pd}$ is always less than 1 even at very high temperatures.
In contrast to the fast diffusers, the slow diffusers exhibit a wide window of temperatures above $T_e$ where $0<D_{pd}<1$.
At low temperatures ($T < T_{drag}$), where $G_V > 0$ and $D_{pd} < 0$, Re and W exhibit an enrichment tendency via the vacancy drag mechanism. Overall, the slow diffusers always exhibit an enrichment tendency regardless of the IKE (i.e., solute-vacancy exchange) or the drag mechanism. 
Note that Co doesn't exhibit a drag regime within the considered temperature range in Figure \ref{fig:drag}; our calculations predicted vacancy drag to become effective only below $T_{drag} = 20 $ K.

Figure~\ref{fig:schematic} summarizes the conditions (temperature, $G_V$, and $D_{pd}$) under which the different solute diffusion mechanisms operate and schematically illustrates the expected solute segregation behaviors near a GB acting as a vacancy sink. Note that when the GB acts as a vacancy source, the directions of component fluxes must be reversed. Hence, under equivalent conditions, the segregation tendencies will be exactly opposite to that discussed in this section. The effect of mechanical strain on $G_V$ and $D_{pd}$, and thereby, on the characteristic temperatures $T_{drag}$ and $T_e$, will be presented in the forthcoming section.


\begin{table}[htp!]
    \caption{Transition temperatures, $T_{drag}$, for the transition from the vacancy drag regime at a low temperature to the IKE regime (no drag) at high temperatures and $T_e$ for crossover in segregation between enrichment and depletion.}
    \centering
    \begin{tabular}{|c|c|c|c|c|c|c|c|}
    \hline
         & Co & Cr & Mo & Re & Ta & W \\
    \hline
       $T_{drag}$ (K) &20 &390 &440 &420 &610 &410 \\
    \hline
    $T_e$ (K)         &--- &440 &680 &--- &620 &--- \\
    \hline
    \end{tabular}
    \label{tab:drag_transition}
\end{table}

\subsection{Strain Effects} \label{sec:strain_effects}
To gain insight into the segregation tendencies of solutes in FCC Ni under mechanical strain, we utilized linear elasticity theory \cite{varvenne_elastic_2017} as implemented in the KineCluE code \cite{schuler_kineclue_2020}. The elastic dipole tensors of various point defects are calculated by DFT, and these are used to estimate the variations in binding and migration energies as a function of strain. The elastic dipole tensors of solute substitutionals and solute-vacancy pairs at the 1nn site are listed in Table \ref{tab:elastic_dipole}; they show the relative amount of residual stress exerted by each species in FCC Ni. The table shows that Ta substitutional exerts the largest compressive stress in FCC Ni (7.9 eV $I$), followed by W (5.5 eV $I$), Mo (5.4 eV $I$), and Re (5.0 eV $I$); here, $I$ is the identity matrix. Furthermore, the Cr substitutional exerts a relatively low residual stress (2.7 eV $I$), while Co does not exert any significant compressive or tensile stress. The transport coefficients under strain are then calculated using the modified strain-dependent defect energetics. The effect of strain on attempt frequencies is neglected since it was shown in an earlier study \cite{garnier_diffusion_2014} for Si in FCC Ni that variations in attempt frequencies are limited to below 10\%. 

The binding $E_b^{VB}[\epsilon_{ij}]$ and migration $E_m^{i\rightarrow f}[\epsilon_{ij}]$ energies under the influence of the strain tensor $\epsilon_{ij}$ are calculated as follows:
\begin{equation}
    E_b^{VB}[\epsilon_{ij}] = E_b^{VB}[0] + P_{ij}^{VB} \epsilon_{ij},
    \label{eq_binding_strain}
\end{equation}
\begin{equation}
    E_m^{i\rightarrow f}[\epsilon_{ij}] = E_m[0] - P_{ij}^{ts, i\rightarrow f} \epsilon_{ij}.
    \label{eq_migration_strain}
\end{equation}
where $P_{ij}^{VB} \epsilon_{ij}$ is the elastic dipole tensor for a solute-vacancy pair and $P_{ij}^{ts, i\rightarrow f}$ is the elastic dipole tensor at the saddle point configuration for a jump between configurations $i$ and $f$.
The sign convention in Equations \ref{eq_binding_strain} and \ref{eq_migration_strain} follow the implementation in KineCluE~\cite{schuler_kineclue_2020}, wherein positive binding energies indicate attraction and negative binding energies indicate repulsion. Moreover, positive stresses in dipole tensors ($P_{ij}$) are compressive and the negative stresses are tensile. However, note that the strain tensors {$\epsilon_{ij}$} use the opposite sign convention, where negative strain is compression and positive strain is tension. 
In this work, we consider the simple case of tensile hydrostatic strain, 
\begin{equation}
    \boldsymbol{\epsilon} = \begin{pmatrix}
    +\epsilon & 0 & 0 \\
    0 & +\epsilon & 0 \\
    0 & 0 & +\epsilon \\
\end{pmatrix}.
\end{equation}

\begin{figure}
    \centering
    \includegraphics[scale = 0.8]{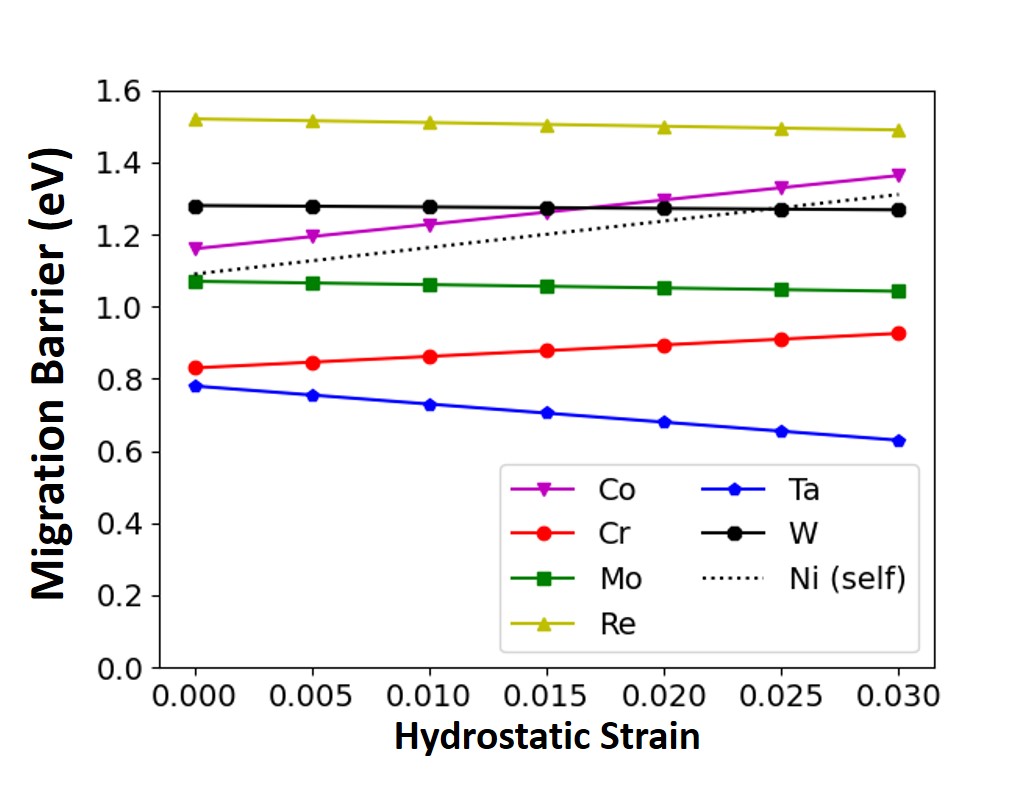}
    \caption{Variation in the migration barriers of solute-vacancy exchange jump with hydrostatic strain. The dotted line indicates the migration barrier for Ni self-diffusion via the vacancy exchange mechanism.}
    \label{fig:strain_dependent_migration_barriers}
\end{figure}

 The strain-dependent barriers calculated using Equation \ref{eq_migration_strain} are plotted in Figure \ref{fig:strain_dependent_migration_barriers}, and the values of  $ P_{ij}^{ts, i\rightarrow f}(X-Vacancy)$ are provided in Table \ref{tab:elastic_dipole}.
 Figure \ref{fig:strain_dependent_migration_barriers} shows that the solute-vacancy exchange barriers for solutes with high residual stresses (Mo, Re, Ta, and W) decrease with tensile hydrostatic strain. In contrast, the solute-vacancy exchange barriers for solutes with low residual stresses (Co and Cr) increase; this is similar to the trend observed for the self-migration barrier of Ni.  
 The variation in the binding and migration barriers of the other configurations and jumps in Figure \ref{fig:fcc_structure} are accounted for, similarly. The corresponding dipole tensors are calculated using DFT but are not reported in this paper.

\begin{figure}[htp!]
    \centering
    \includegraphics[width=\textwidth]{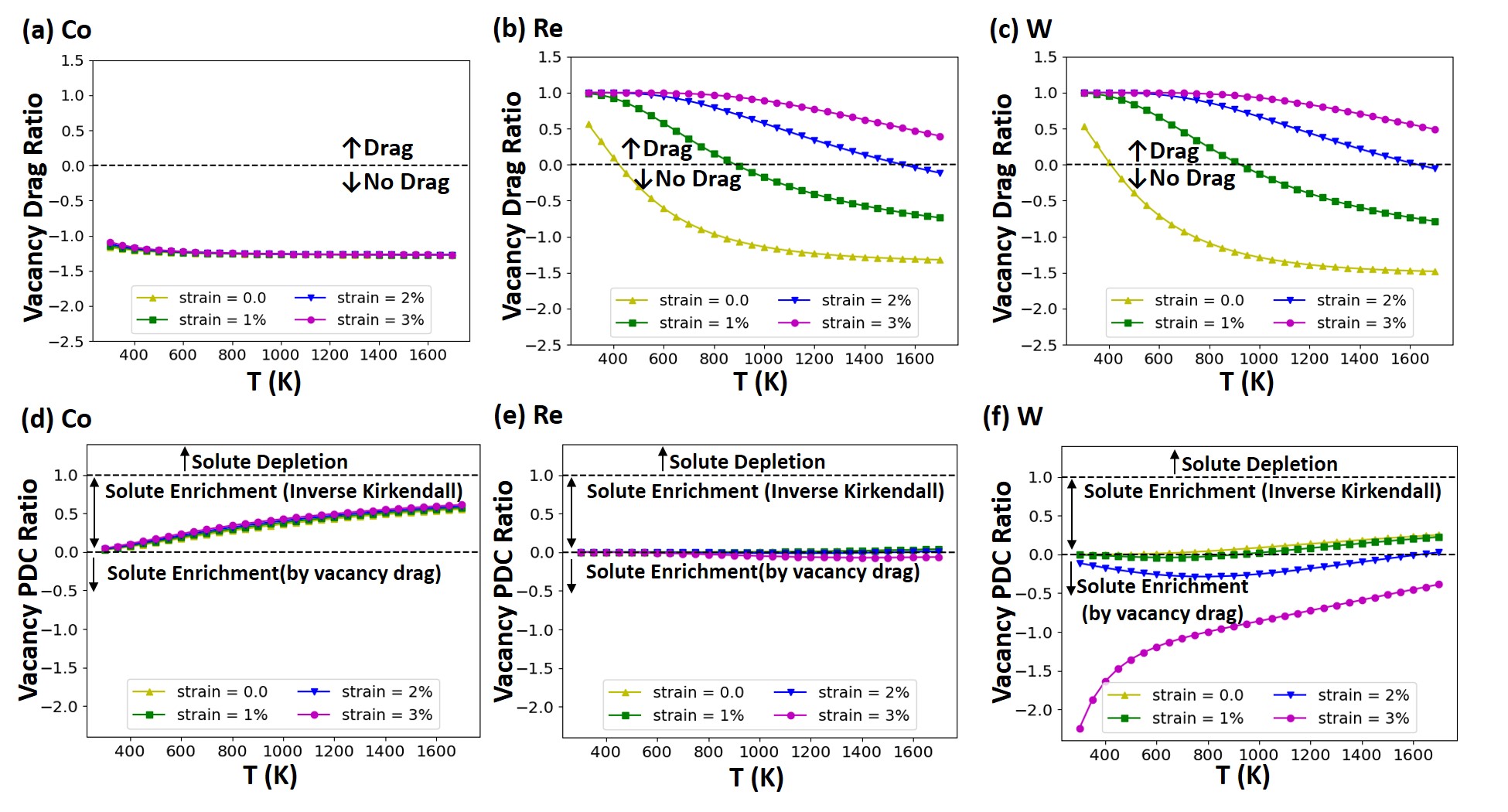}
    \caption{Vacancy drag ratios of (a) Co, (b) Re, and (c) W in FCC Ni under a hydrostatic strain of 0\%, 1\%, 2\%, and 3\%. The partial diffusion coefficient ratios of these solutes under the same strain values are plotted in panels (d), (e), and (f), respectively, on a semilogarithmic plot.}
    \label{fig:strain_slow_diff}
\end{figure}

To understand the solute segregation tendencies under the influence of tensile hydrostatic strain, we first analyze the vacancy drag $G_V$ and partial diffusion coefficient $D_{pd}$ ratios of the slow diffusing solutes (Co, Re, and W) plotted in Figure \ref{fig:strain_slow_diff}. Tensile hydrostatic strains of 1\%, 2\%, and 3\% were considered. For Co, Figures \ref{fig:strain_slow_diff}a and \ref{fig:strain_slow_diff}d show that both the vacancy drag mechanism and the segregation tendency are not affected by the strain. This is because Co substitution in FCC Ni introduces a zero residual stress in the matrix. Moreover, the elastic dipole tensor of the Co-vacancy pair is almost identical to that of a monovacancy (see Table \ref{tab:elastic_dipole}). In other words, a Co atom in FCC Ni acts like a solvent atom, except that it is slightly slower to diffuse. Consequently, in the presence or absence of strain, Co is not dragged by vacancies. It is expected to enrich at vacancy sinks through the inverse Kirkendall mechanism due to its slower diffusivity relative to Ni. By comparing the similar variation in the migration energies of Co and Ni under strain (Figure \ref{fig:strain_dependent_migration_barriers}), we infer the mobility of Co to vary by almost the same amount as that of Ni. 

\begin{table}[htp!]
    \centering
    \caption{The DFT calculated elastic dipole tensors for substitutional ( ($P_{ij}(X)$)), substitutional-vacancy pair at the $1^{st}$ nn site ($P_{ij}(X-Vacancy)$) defects, and for the saddle point of substitutional-vacancy exchange jump ( ($ P_{ij}^{ts, i\rightarrow f}(X-Vacancy)$)) in FCC Ni. The substitutional vacancy pairs are aligned in the [011] crystallographic direction. Positive values represent a compressive stress state, while negative values represent a tensile stress state. }
    \begin{tabular}{|c|c|c|c|}
    \hline
        X & $P_{ij}(X)$  (eV) & $P_{ij}(X-Vacancy)$ (eV) & $ P_{ij}^{ts, i\rightarrow f}(X-Vacancy)$ (eV) \\
        \hline
       Ni & --- &  $\begin{pmatrix}
           -4.6 &  0.0   & 0.0 \\
            0.0   & -4.6 & 0.0 \\
            0.0   & 0.0 & -4.6 \\
       \end{pmatrix}$ &$\begin{pmatrix}
           2.8 &  0.0  &  0.0 \\
           0.0 & -5.1  &  0.1 \\
           0.0 &  0.1  & -5.1 \\
       \end{pmatrix}$ \\
        \hline
       Co &$\begin{pmatrix}
            0.0 &  0.0   & 0.0 \\
            0.0   & 0.0 & 0.0 \\
            0.0   & 0.0 & 0.0 \\
       \end{pmatrix}$ & $\begin{pmatrix}
           -4.7 &  0.0 & 0.0 \\
            0.0 & -4.6 & 0.1 \\
            0.0 & 0.1  & -4.6 \\
       \end{pmatrix}$ & $\begin{pmatrix}
           2.6 &  0.0  &  0.0 \\
           0.0 & -4.7  &  0.4 \\
           0.0 &  0.4  & -4.7 \\
       \end{pmatrix}$ \\
       \hline
       Cr &$\begin{pmatrix}
            2.7 &  0.0   & 0.0 \\
            0.0   & 2.7 & 0.0 \\
            0.0   & 0.0 & 2.7 \\
       \end{pmatrix}$ &$\begin{pmatrix}
            -1.5 &  0.0   & 0.0 \\
            0.0   & -2.5 & 0.6 \\
            0.0   & 0.6 & -2.5 \\
       \end{pmatrix}$ &$\begin{pmatrix}
           3.9 &  0.0  &  0.0 \\
           0.0 & -3.5  &  1.7 \\
           0.0 &  1.7  & -3.5 \\
       \end{pmatrix}$  \\
       \hline
       Mo &$\begin{pmatrix}
            5.4 &  0.0   & 0.0 \\
            0.0   & 5.4 & 0.0 \\
            0.0   & 0.0 & 5.4 \\
       \end{pmatrix}$ &$\begin{pmatrix}
            0.9 &  0.0   & 0.0 \\
            0.0   & -0.5 & 1.1 \\
            0.0   & 1.1 & -0.5 \\
       \end{pmatrix}$ &$\begin{pmatrix}
           5.5 &  0.0  &  0.0 \\
           0.0 & -2.3  &  2.0 \\
           0.0 &  2.0  & -2.3 \\
       \end{pmatrix}$  \\
       \hline
       Re &$\begin{pmatrix}
            5.0 &  0.0   & 0.0 \\
            0.0   & 5.0 & 0.0 \\
            0.0   & 0.0 & 5.0 \\
       \end{pmatrix}$ &$\begin{pmatrix}
           0.0 &  0.0 & 0.0 \\
            0.0 & -0.5 & 0.8 \\
            0.0 & 0.8  & -0.5 \\
       \end{pmatrix}$ &$\begin{pmatrix}
           6.4 &  0.0  &  0.0 \\
           0.0 & -2.7  &  1.2 \\
           0.0 &  1.2  & -2.7 \\
       \end{pmatrix}$  \\
       \hline
       Ta &$\begin{pmatrix}
            7.9 &  0.0   & 0.0 \\
            0.0   & 7.9 & 0.0 \\
            0.0   & 0.0 & 7.9 \\
       \end{pmatrix}$ &$\begin{pmatrix}
            3.8 &  0.0   & 0.0 \\
            0.0   & 1.3 & 1.4 \\
            0.0   & 1.4 & 1.3 \\
       \end{pmatrix}$ &$\begin{pmatrix}
           6.8 &  0.0  &  0.0 \\
           0.0 & -0.9  &  2.8 \\
           0.0 &  2.8  & -0.9 \\
       \end{pmatrix}$   \\
       \hline
       W &$\begin{pmatrix}
            5.5 &  0.0   & 0.0 \\
            0.0   & 5.5 & 0.0 \\
            0.0   & 0.0 & 5.5 \\
       \end{pmatrix}$ &$\begin{pmatrix}
            0.9 &  0.0   & 0.0 \\
            0.0   & -0.3 & 0.9 \\
            0.0   & 0.9 & -0.3 \\
       \end{pmatrix}$ &$\begin{pmatrix}
           5.9 &  0.0  &  0.0 \\
           0.0 & -2.8  &  1.4 \\
           0.0 &  1.4  & -2.8 \\
       \end{pmatrix}$   \\
       \hline
        
    \end{tabular}
    \label{tab:elastic_dipole}
\end{table}

The other two slow diffusers, Re and W, exert considerable residual stress in FCC Ni (Table \ref{tab:elastic_dipole}). As a result, they experience a significant change in the vacancy drag behavior ($G_V$ in Figures \ref{fig:strain_slow_diff}b and \ref{fig:strain_slow_diff}c) in the presence of strain, with $T_{drag}$ being shifted to higher temperatures. At 3\% strain, both Re and W exhibit positive $G_V$ for all temperatures between 300 and 1700 K. Despite the significant effect of strain on $G_V$, $D_{pd}$ and the segregation tendencies are observed to be weakly affected. For instance, Re---the slowest diffuser---shows a negligible change in $D_{pd}$ even at a strain of 3\% (Figure \ref{fig:strain_slow_diff}e).  
This is because of the relatively high migration barriers of these solutes.
Under both strained and unstrained conditions, the diffusivity of Re relative to Ni is negligible (i.e., $D_{pd} \approx 0$). While adding a tensile hydrostatic strain of 3\% changes the direction of the Re diffusion due to the strong binding with vacancy, it does not change the Re enrichment tendency; it only switches the enrichment mechanism from the inverse Kirkendall to vacancy drag as $D_{pd}$ changes sign from a positive to a negative value. Since W is slightly faster than Re, a more noticeable change in the magnitude of $D_{pd}$ is observed for strains of 2\% and 3\% (Figure \ref{fig:strain_slow_diff}e). 

\begin{figure}[htp!]
    \centering
    \includegraphics[width=\textwidth]{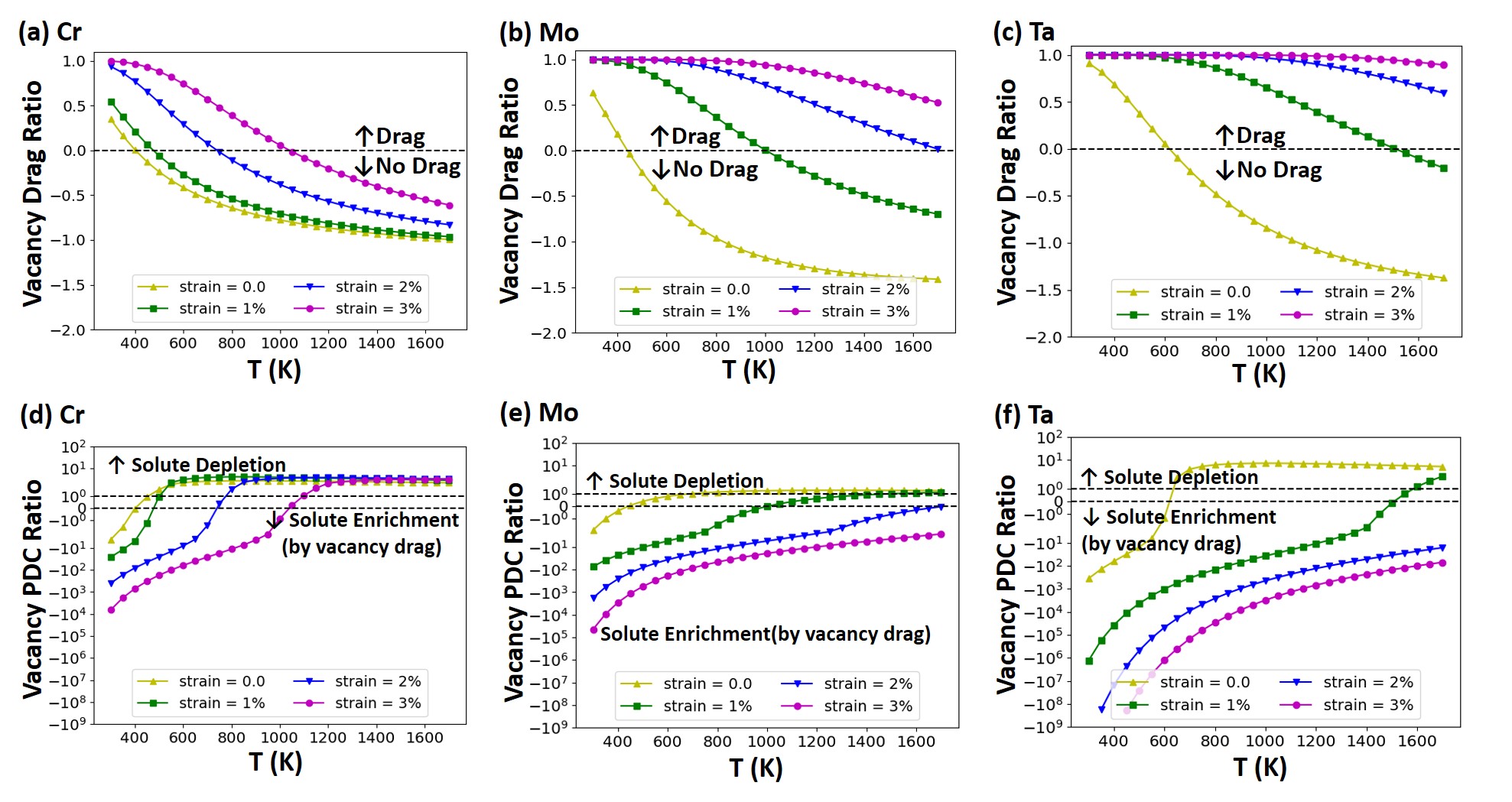}
    \caption{Vacancy drag ratios of (a)Cr, (b)Mo, and (c)Ta in FCC Ni under a tensile hydrostatic strain of 0\%, 1\%, 2\%, and 3\%. The partial diffusion coefficient ratios of these three solutes under the same strain values are plotted in panels (d), (e), and (f), respectively, on a semilogarithmic scale.}
    \label{fig:strain_fast_diff}
\end{figure}

We now analyze the vacancy drag $G_V$ and partial diffusion coefficient ratios $D_{pd}$ of the fast diffusing solutes (Cr, Mo, and Ta) plotted in Figure \ref{fig:strain_fast_diff}.
In contrast to the slow-diffusing solutes, $D_{pd}$ and the segregation tendencies of the fast-diffusing solutes, Cr, Mo, and Ta, are significantly affected by strain, as shown in Figure \ref{fig:strain_fast_diff}. The
$G_V$ of Cr changes moderately with increasing hydrostatic strain (Figure \ref{fig:strain_fast_diff}a). It is not as significant of a dependency compared to solutes (Re, W, Mo, and Ta) that exert a large compressive residual stress; at the same time, it is not negligible as in the case of Co because of its nonzero dipole tensor (Table \ref{tab:elastic_dipole}). 
On the other hand, the change in $G_V$ of Mo and Ta is dramatic (Figures \ref{fig:strain_fast_diff}b and \ref{fig:strain_fast_diff}c), which is similar to that of Re and W. However, in contrast to Re and W, the effect of strain for Mo and Ta reflects a dramatic change in $D_{pd}$ (Figures \ref{fig:strain_fast_diff}b and \ref{fig:strain_fast_diff}c). 
The effect is more pronounced for Ta; for instance, at T = 700 K, a hydrostatic strain of 1\% leads to the decrease in $D_{pd}$ from 3.85 (depletion tendency due to inverse Kirkendall) to -3.4$\times 10^2$ (strong enrichment tendency due to vacancy drag). A strain of 3\% at the same temperature decreases $D_{pd}$ further by three orders of magnitude.

In summary, our results indicate that the segregation tendency of the fast-diffusing solutes is highly sensitive to strain and that tensile hydrostatic strain tends to alter their segregation tendency from depletion to enrichment.
For these solutes, tensile hydrostatic strain causes a shift in both $T_{drag}$ and $T_e$ to higher temperatures.
While strain alters the direction of the flux of the slow-diffusing solutes relative to that of vacancies (especially for solutes with larger atomic radii, such as Re and W), it does not alter the segregation tendency because they already exhibit an enrichment tendency due to their low mobility.
For these solutes (except for Co), tensile hydrostatic strain causes a shift in $T_{drag}$ to higher temperatures.

\subsection{Relevance to Diffusional Creep}
When a polycrystalline material is subjected to an applied tensile stress, GBs parallel and perpendicular to the applied stress direction experience compressive and tensile stress fields, respectively, around them \cite{villani2015field, RN475}. As a result, the parallel GBs will have a reduced concentration of equilibrium vacancies, while the perpendicular GBs will have an enhanced concentration of the same. As such, under diffusional creep conditions, the N-H theory premises that a net flux of vacancies diffuses towards the parallel GBs, to be absorbed (i.e., the parallel GBs act as vacancy sinks). In contrast, the perpendicular GBs (under tension) act as vacancy sources and receive a net atomic flux. 

Since diffusional creep in Ni-based alloys is deemed significant at temperatures above 750 K, the pertinent temperatures of interest are well above $T_{drag}$. 
Without considering strain effects, the segregation tendencies of the various solutes, predicted for vacancy sinks in Section~\ref{sec:drag_segregation}, are relevant to the parallel GBs.
Our analysis points out that the solute-vacancy exchange mechanism and IKE is dominant at the high temperatures under which the diffusional creep mechanism is active. Furthermore, depletion of the fast diffusers (Cr, Mo, and Ta) and enrichment of the slow diffusers (Co, Re, and W) are expected at the parallel GBs.
An opposite segregation tendency is expected for these solutes at the perpendicular GBs since these GBs act as vacancy sources and the direction of fluxes (see Figure~\ref{fig:schematic}) are reversed.

Considering strain effects, the results of Section~\ref{sec:strain_effects}, which pertain to tensile strain, are relevant to the perpendicular boundaries. However, the segregation tendencies reported therein must be reversed since the results were interpreted for segregation at vacancy sinks, whereas perpendicular boundaries act as vacancy sources.
For parallel GBs (vacancy sinks), the effects of compressive strain are more relevant. Without a loss of generality, we may extrapolate the trends in $G_V$ and $D_{pd}$ obtained for tensile strains towards compressive strains.
Such an extrapolation of our results indicates that compressive strain in the vicinity of parallel GBs will lower $T_{drag}$, thus making IKE the dominant mechanism. This is expected to strengthen the solute segregation tendencies predicted at these GBs in the absence of strain. Hence, a stronger depletion of Cr, Mo, and Ta and a stronger enrichment of Re and W are expected. Additionally, strain effects are expected to have little influence on the segregation tendency of Co. The results provide us a basis for choosing suitable alloy systems and the expected solute segregation tendencies at GBs for our future experimental investigations to verify the occurrence of diffusional creep mechanisms. 
{In a recent mesoscale modeling study, Hussein et al.\mbox{~\cite{hussein2024effect}} showed segregation of elements in Fe-Au alloy under creep conditions. The results of their diffusion model coupled with polycrystal elastic stress fields demonstrated that the segregation tendency is affected by the type and extent of stress field around GBs. For instance, Au, the faster solute, was predicted to enrich at the GBs under tensile stress. Our results and methodology enable more accurate parameterization of such mesoscale models.}


\section{Conclusions}
In this work, solute-vacancy interactions and transport properties of Co, Cr, Mo, Re, Ta, and W in FCC Ni were obtained using DFT calculations. We employed the DFT energetics to evaluate the Onsager transport coefficients using the SCMF implemented in the KineCluE code. Solute diffusivities calculated using this approach showed good agreement with prior theoretical calculations and experimental tracer diffusivity measurements in the literature. Furthermore, using vacancy drag and partial diffusion coefficient ratios evaluated from the Onsager transport coefficients, we analyzed the diffusion mechanisms and segregation tendencies as a function of temperature and tensile hydrostatic strain.

We found Co, Re, and W to be slow diffusers at high temperatures, while Cr, Mo, and Ta to be fast diffusers. Due to the favorable solute-vacancy exchange at high temperatures (above 440 K for Cr, 680 K for Mo, and 620 K for Ta), the fast diffusers were found to exhibit a depletion tendency (IKE) at vacancy sinks.
However, as the vacancy drag mechanism became dominant at lower temperatures,
the fast diffusers demonstrated a segregation transition to enrichment. 
In contrast, due to the unfavorable solute-vacancy exchange (at temperatures above 20 K for Co, 420 K for Re, and 410 K for W), the slow diffusers were found to exhibit an enrichment tendency at vacancy sinks even at higher temperatures. 
While the vacancy drag mechanism became dominant at lower temperatures, thus changing their diffusion mechanism, their segregation tendency remained as enrichment.

Under the assumption of linear elasticity and hydrostatic strain, we identified the following two properties to control the sensitivity of segregation tendency to strain:
\vspace{-\topsep}
\begin{enumerate}
\setlength{\parskip}{0pt}
\setlength{\itemsep}{0pt plus 1pt}
\item The residual stress introduced by the solute atom in the matrix:
When subjected to strain, the vacancy drag ratios of solutes (Re, W, Mo, and Ta) that imparted a high residual stress in FCC Ni changed more significantly than those that imparted a very low residual stress (Cr and Co). Solutes imparting high residual stress were found to cause a considerable change in segregation tendency when they were also fast diffusers.
\item The mobility of the solute relative to the solvent:
The segregation tendency (i.e., the numerical value of the partial diffusion coefficient ratio $D_{pd}$) changed more significantly for fast diffusers (Mo, Cr, and Ta), from depletion under strain-free condition to enrichment under tensile strain. In contrast, the change in $D_{pd}$ was minimal for slow diffusers (Co, Re, and W). 
\end{enumerate}
\vspace{-\topsep}
Finally, the implications of our results on the segregation tendencies at GBs parallel and perpendicular to the applied stress direction, in polycrystals subject to diffusional creep conditions, were discussed.

Our theoretical study indicates that the N-H diffusional creep mechanism is likely to induce nonequilibrium and nonuniform solute segregation at GBs in FCC Ni-based binary alloys. At the high temperatures relevant to diffusional creep, the IKE is expected to dominate, and thus, fast diffusers are likely to deplete and slow diffusers are likely to enrich at GBs parallel to the applied tensile stress. Our conclusions further support the use of slow-diffusing solutes such as Re (which demonstrated a stable enrichment tendency versus temperature and strain) in Ni-based superalloys for high-temperature creep applications. The Onsager transport coefficients derived in this work are expected to serve as input to mesoscale microstructure-level models to enable a more rigorous assessment of the solute segregation behavior under N-H creep conditions.
 
\appendix

\section{Partition Functions and Cluster Fractions}
\label{appendix_partition}

In this appendix, we present details on how the total defect concentration $C$ and the cluster fractions $f_V$ and $f_{VB}$ in Equation \ref{eq_onsager_matrix} are evaluated. $C$ is the sum of concentration of the monovacancies $C_V$ and solute-vacancy pairs $C_{VB}$ and is calculated as follows:
\begin{equation}
\label{eq_tot_conc}
    C = C_V + C_{VB} - C_{corr},
\end{equation}
where $C_{corr}$ is a correction term introduced by Messina et al.~\cite{messina_solute_2020} to account for the sites that single vacancies cannot occupy because of the geometrical definition of the solute-vacancy pairs.
The concentration terms,$C_V$, $C_{VB}$, and $C_{corr}$, are calculated according to the following equations:
\begin{equation}
\label{eq_vac_conc}
    C_V = Z_V [V],
\end{equation}
\begin{equation}
    C_{VB} = Z_{VB} [V] [B],
\end{equation}
\begin{equation}
\label{eq_corr_conc}
    C_{corr} = Z_{VB}^0 [V] [B].
\end{equation}

First, we define the partition function terms $Z_V$, $Z_{VB}$, and $Z_{VB}^0$. $Z_V$ is the monovacancies partition function and is equal to the number of symmetrically equivalent configurations in the FCC crystal structure, which is 1. $Z_{VB}$ is the solute-vacancy pair partition function and is evaluated in KineCluE by summation of $exp(\frac{E_{bind}^i}{k_B T})$ for all possible solute-vacancy pair configurations $i$ in the cluster~\cite{schuler_kineclue_2020}. 
The term $Z_{VB}^0$ is the noninteracting cluster partition function and is equal to the number of possible pair configurations within the specified kinetic radius ($r_{kin}$). KineCluE evaluates all the possible jumps up to a maximum trajectory that is equal to $r_{kin}$ and also calculates $Z_{VB}^0$ for that $r_{kin}$. Larger $r_{kin}$ values mean accounting for more long-range kinetic correlations and hence providing more accurate transport coefficients. 

In this work, we performed a convergence test using $r_{kin}$ values of $2a, 3a, 4a,$ and $5a$ (where ``a" is the lattice constant of FCC Ni). A value of $4a$ achieved good convergence and therefore was used in all calculations. This resulted in $Z_{VB}^0 = 1060$. Also, the maximum solute concentration $[B]_{max}$ allowed in this dilute model can be derived from the fact that the $C_{corr}$ value cannot exceed the value of $C_V$. By applying this condition using  Equations (\ref{eq_vac_conc}) and (\ref{eq_corr_conc}):
\begin{equation}
    Z_{VB}^0 [V] [B] < Z_V [V] \Longrightarrow [B] < \frac{Z_V}{Z_{VB}^0},
\end{equation}
and the maximum solute concentration in the dilute limit will be:
\begin{equation}
        [B]_{max} = \frac{1}{1060} = 0.094\%.
\end{equation}
Then, we define $[V]$, which is the Boltzmann expression for the thermodynamic equilibrium concentration of vacancies in the dilute limit:
\begin{equation}
    [V] = e^{\frac{-E_{form}^{vac}}{k_B T}} e^{\frac{S_{form}^{vac}}{k_B}}
\end{equation}
where $E_{form}^{vac}$ is the vacancy formation energy  and $S_{form}^{vac}$ is the vacancy formation entropy reported in Table \ref{tab:bulk_prop}. 
Finally, the cluster fractions are defined by:
\begin{equation}
    f_V = \frac{C_V - C_{corr}}{C},
\end{equation}
\begin{equation}
    f_{VB} = \frac{C_{VB}}{C}.
\end{equation}

\section{DFT Data}
\label{appendix_DFT_data}
\begin{table}
    \caption{Solute-vacancy binding energies in FCC nickel. Positive binding energies indicate attraction while negative binding energies indicate repulsion.}
    \centering    
    \begin{tabular}{|c|c|c|c|c|c|c|}
        \hline
         & Cr-vacancy & Re-vacancy & Ta-vacancy & W-vacancy & Mo-vacancy & Co-vacancy \\
         \hline
         BE$_{1nn}$(eV)  &-0.047 &-0.040 & 0.080 &-0.020 & 0.008 &-0.024 \\  
         BE$_{2nn}$(eV)  & 0.002 & 0.068 &-0.029 & 0.023 & 0.021 & 0.011 \\
         BE$_{3nn}$(eV)  &-0.032 &-0.034 &-0.025 &-0.034 &-0.030 &-0.021 \\
         BE$_{4nn}$(eV)  & 0.043 & 0.048 & 0.086 & 0.064 & 0.065 & 0.010 \\
         BE$_{5nn}$(eV)  &-0.013 &-0.008 &-0.010 &-0.011 &-0.010 &-0.005 \\
         BE$_{6nn}$(eV)  & 0.003 & 0.015 & 0.010 & 0.011 & 0.011 &-0.003 \\
         \hline
    \end{tabular}
    \label{tab:binding_energies}
\end{table}

Table \ref{tab:binding_energies} shows the binding energies for solute-vacancy pairs at separation distances up to the sixth-nearest neighbor. 

Table \ref{tab:migration_energies} shows the migration energies for all the vacancy jumps considered in this work. The DFT barriers calculated for Ni-Cr by Toijer et al. \cite{toijer_solute-point_2021} and for Ni-Re, Ni-Ta, Ni-W, and Ni-Mo by Schuwalow et al. \cite{schuwalow_vacancy_2014} are also presented in Table \ref{tab:migration_energies} for comparison. It should be noted that the five-frequency model adopted by Toijer et al. \cite{toijer_solute-point_2021} does not require the calculation of the migration barriers beyond the first-nearest neighbor (2nn$\rightarrow$3nn, 3nn$\rightarrow$3nn, and 3nn$\rightarrow$4nn). However, we extend our model to include all possible jumps within the fourth-nearest neighbor shell. This is a more conservative treatment to account for the effect of solute-vacancy interactions on vacancy migration.

\begin{table}
    \caption{Vacancy migration barriers in eV for atomic the vacancy jumps in Ni-X systems. X: Cr, Re, Ta, W, Mo, Co. Values from previous DFT studies in the literature~\cite{toijer_solute-point_2021, schuwalow_vacancy_2014} are shown in parentheses for comparison.}
    \label{tab:migration_energies}
    \centering
    \begin{tabular}{|c|c|c|c|c|c|c|}
    \hline
        Jump & Ni-Cr & Ni-Re & Ni-Ta & Ni-W & Ni-Mo & Ni-Co  \\
        \hline
        Solute-vacancy exchange &0.83 &1.52 &0.78 &1.28 &1.07 &1.16 \\
         &(0.77 \cite{toijer_solute-point_2021}) & (1.46 \cite{schuwalow_vacancy_2014}) &(0.75 \cite{schuwalow_vacancy_2014}) &(1.22 \cite{schuwalow_vacancy_2014}) &(1.03 \cite{schuwalow_vacancy_2014}) & \\
        \hline
        1nn$\rightarrow$1nn     &0.99 &1.14 &1.39 &1.22 &1.22 &1.08 \\
         &(1.04 \cite{toijer_solute-point_2021}) &(1.05 \cite{schuwalow_vacancy_2014}) &(1.33 \cite{schuwalow_vacancy_2014}) &(1.16 \cite{schuwalow_vacancy_2014}) &(1.16 \cite{schuwalow_vacancy_2014}) & \\
        \hline
        1nn$\rightarrow$2nn     &1.03 &0.96 &1.02 &0.98 &0.99 &1.08 \\
         &(1.09 \cite{toijer_solute-point_2021}) &(0.89 \cite{schuwalow_vacancy_2014}) &(0.96 \cite{schuwalow_vacancy_2014}) &(0.91 \cite{schuwalow_vacancy_2014}) &(0.93 \cite{schuwalow_vacancy_2014}) & \\
        \hline
        2nn$\rightarrow$1nn     &1.08 &1.06 &0.90 &1.01 &0.99 &1.12 \\
         &(1.14 \cite{toijer_solute-point_2021}) &(0.98 \cite{schuwalow_vacancy_2014}) &(0.83 \cite{schuwalow_vacancy_2014}) &(0.93  \cite{schuwalow_vacancy_2014}) &(0.91 \cite{schuwalow_vacancy_2014}) & \\
        \hline
        1nn$\rightarrow$3nn     &1.05 &1.08 &1.02 &1.06 &1.05 &1.10 \\
         &(1.12 \cite{toijer_solute-point_2021}) &(1.00 \cite{schuwalow_vacancy_2014}) &(0.96 \cite{schuwalow_vacancy_2014}) &(0.98 \cite{schuwalow_vacancy_2014}) &(0.98 \cite{schuwalow_vacancy_2014}) & \\
        \hline
        3nn$\rightarrow$1nn     &1.06 &1.09 &0.92 &1.05 &1.01 &1.11 \\
         &(1.13 \cite{toijer_solute-point_2021}) &(1.02 \cite{schuwalow_vacancy_2014}) &(0.86 \cite{schuwalow_vacancy_2014}) &(0.98 \cite{schuwalow_vacancy_2014}) &(0.95 \cite{schuwalow_vacancy_2014}) & \\
        \hline
        1nn$\rightarrow$4nn     &1.05 &1.07 &1.04 &1.06 &1.05 &1.09 \\
         &(1.11 \cite{toijer_solute-point_2021}) &(1.00 \cite{schuwalow_vacancy_2014}) &(0.98 \cite{schuwalow_vacancy_2014}) &(0.99 \cite{schuwalow_vacancy_2014}) &(0.99 \cite{schuwalow_vacancy_2014}) & \\
        \hline
        4nn$\rightarrow$1nn     &1.14 &1.16 &1.04 &1.14 &1.10 &1.13 \\
         &(1.17 \cite{toijer_solute-point_2021}) &(1.09 \cite{schuwalow_vacancy_2014}) &(0.98 \cite{schuwalow_vacancy_2014}) &(1.07 \cite{schuwalow_vacancy_2014}) &(1.04 \cite{schuwalow_vacancy_2014}) & \\
        \hline
        2nn$\rightarrow$3nn     &1.10 &1.14 &1.13 &1.13 &1.12 &1.10 \\
         & &(1.05 \cite{schuwalow_vacancy_2014}) &(1.05 \cite{schuwalow_vacancy_2014}) &(1.05 \cite{schuwalow_vacancy_2014}) &(1.04 \cite{schuwalow_vacancy_2014}) & \\
        \hline
        3nn$\rightarrow$2nn     &1.07 &1.05 &1.15 &1.09 &1.09 &1.07 \\
         & &(0.97 \cite{schuwalow_vacancy_2014}) &(1.09 \cite{schuwalow_vacancy_2014}) &(1.03 \cite{schuwalow_vacancy_2014}) &(1.02 \cite{schuwalow_vacancy_2014}) & \\
        \hline
        3nn$\rightarrow$3nn     &1.08 &1.08 &1.07 &1.08 &1.07& \\
         & &(0.99 \cite{schuwalow_vacancy_2014}) &(1.00 \cite{schuwalow_vacancy_2014}) &(1.00 \cite{schuwalow_vacancy_2014}) &(1.00 \cite{schuwalow_vacancy_2014}) & \\
        \hline
        3nn$\rightarrow$4nn     &1.09 &1.09 &1.11 &1.09 &1.09 &1.08 \\
         & &(1.01 \cite{schuwalow_vacancy_2014}) &(1.05 \cite{schuwalow_vacancy_2014}) &(1.02 \cite{schuwalow_vacancy_2014}) &(1.02 \cite{schuwalow_vacancy_2014}) & \\
        \hline
        4nn$\rightarrow$3nn     &1.16 &1.16 &1.21 &1.17 &1.17 &1.11 \\
         & &(1.08 \cite{schuwalow_vacancy_2014}) &(1.14 \cite{schuwalow_vacancy_2014}) &(1.11 \cite{schuwalow_vacancy_2014}) &(1.10 \cite{schuwalow_vacancy_2014}) & \\
         
        \hline
    \end{tabular}
\end{table}

\FloatBarrier

\section*{ACKNOWLEDGMENTS}

The work was supported by the U.S. Department of Energy, Office of Science, Materials Sciences and Engineering Division. 
The authors thank Daniel Schwen and Jia-Hong Ke of Idaho National Laboratory for reviewing the manuscript and providing helpful feedback.
SS and SBK thank Luca Messina of the French Alternative Energies and Atomic Energy Commission (CEA Cadarache) for helpful discussions on the KineCluE code.  
This research made use of Idaho National Laboratory's High Performance Computing systems located at the Collaborative Computing Center and supported by the Office of Nuclear Energy of the U.S. Department of Energy and the Nuclear Science User Facilities under Contract No. DE-AC07-05ID14517. 

\bibliography{main}

\end{document}